\documentclass{aa}  

\usepackage{graphicx}
\usepackage{txfonts}
\usepackage[colorlinks=true
  ,urlcolor=blue
  ,anchorcolor=blue
  ,citecolor=blue
  ,filecolor=blue
  ,linkcolor=blue
  ,menucolor=blue
  ,linktocpage=true
  ,pdfproducer=medialab
  ,pdfa=true
]{hyperref}
\usepackage{graphicx}	
\usepackage{amsmath}	
\usepackage{amssymb}	
\usepackage{float}
\usepackage[table,xcdraw,dvipsnames]{xcolor}
\usepackage{booktabs}
\usepackage{xspace}
\usepackage{comment}

\newcommand{\pngsim}[0]{\textsc{PNG-UNITsims}\xspace}
\newcommand{\fnl}[0]{$f_{\rm NL}$\xspace}
\newcommand{\bp}[0]{$b_{\phi}$\xspace}

\begin{document}

   \title{PNG-UNITsims: Halo clustering response to primordial non-Gaussianities as a function of mass}
   \author{Adrián Gutiérrez Adame \inst{1,2,3} \thanks{adrian.gutierrez@uam.es} 
          \and Santiago Avila \inst{4,1,3} 
          \and  Violeta Gonzalez-Perez \inst{1,2} 
          \and Gustavo Yepes \inst{1,2} 
          \and  Marcos Pellejero \inst{5,6} 
          \and Mike S. Wang \inst{5}
          \and Chia-Hsun Chuang \inst{7,8} 
          \and Yu Feng \inst{9} 
          \and Juan Garcia-Bellido \inst{1,3} 
          \and Alexander Knebe \inst{1,2,10} 
          }

   \institute{Departamento de F\'isica Te\'orica,  Universidad Aut\'onoma de Madrid, 28049 Madrid, Spain
         \and
             Centro de Investigaci\'on Avanzada en F\'isica Fundamental (CIAFF), Facultad de Ciencias, Universidad Aut\'onoma de Madrid, ES-28049 Madrid, Spain
         \and 
            Instituto de F\'isica Teorica UAM-CSIC, c/ Nicolás Cabrera 13-15, Cantoblanco, 28049 Madrid, Spain
         \and
            Institut de física d’altes energies (IFAE) The Barcelona Institute of Science and Technology campus UAB, 08193 Bellaterra Barcelona, Spain
        \and 
            Institute for Astronomy, University of Edinburgh, Royal Observatory Edinburgh, Blackford Hill, Edinburgh EH9 3HJ, United Kingdom
        \and 
            Donostia International Physics Centre, Paseo Manuel de Lardizabal 4, 20018 Donostia-San Sebastian, Spain.
        \and 
            Department of Physics and Astronomy, The University of Utah, 115 South 1400 East, Salt Lake City, UT 84112, USA
        \and 
             Kavli Institute for Particle Astrophysics and Cosmology, Stanford University, 452 Lomita Mall, Stanford, CA 94305, USA.
        \and 
            Berkeley Center for Cosmological Physics, University of California, Berkeley, CA 94720, USA
        \and
            International Centre for Radio Astronomy Research, University of Western Australia, 35 Stirling Highway, Crawley, Western Australia 6009, Australia
             }

\titlerunning{PNG-UNITsims}
\authorrunning{A.G. Adame et al.}

   \date{Submitted 20 December 2023}


  \abstract{This paper presents the \pngsim suite, which includes the largest full N-body simulation to date with local primordial non-Gaussianities (local PNG), the \textsc{PNG-UNIT}. The amplitude of the PNGs is given by $f_{\rm NL}^{local}=100$. The simulation follows the evolution of $4096^3$ particles in a periodic box with $L_{\rm box} = 1 \; h^{-1}\,{\rm Gpc}$, resulting in a mass resolution of $m_{p} = 1.24\times 10^{9}\; h^{-1}\,M_\odot$, enough to finely resolve the galaxies targeted by stage-IV spectroscopic surveys. The \textsc{PNG-UNIT} has fixed initial conditions with phases also matching the pre-existing \textsc{UNIT} simulation with Gaussian initial conditions. The fixed and matched initial conditions reduce the simulation uncertainty significantly. In this first study of the \pngsim, we measure the PNG response parameter, $p$, as a function of the halo mass. halos with masses between $1\times 10^{12}$ and $5\times 10^{13} \, h^{-1} M_\odot$ are well described by the universality relation, given by $p=1$. For halos with masses between $2\times 10^{10}$ and $1\times 10^{12} \, h^{-1} M_\odot$ we find that $p<1$, at a significance between $1.5$ and $3.1\sigma$. Combining all the halos between $2\times 10^{10}$ and $5\times 10^{13} \, h^{-1} M_\odot$,  we find $p$ consistent with a value of $0.955\pm0.013$, which is $3\sigma$ away from the universality relation. We demonstrate that these findings are robust to mass resolution, scale cuts and uncertainty estimation. We also compare our measurements to separate universe simulations, finding that the \pngsim constraints outperform the former for the setup considered. Using a prior on $p$ as tight as the one reported here for DESI-like forecast can result in \fnl constraints comparable to fixing $p$. At the same time, fixing $p$ to a wrong value $(p=1)$ may result in up to $2\sigma$ biases on \fnl.
  }

   \maketitle
%

\section{Introduction}
Primordial non-Gaussianities (PNG) provide a unique window into the Universe's dynamics during the inflationary era. One of the most widely studied models is that of local primordial non-Gaussianities, where the amplitude of the deviation from Gaussianity is typically parameterised using $f_{\rm NL}^{local}$ \citep{Komatsu_2001}. In this paper, we only consider the local PNG, so we refer to this quantity simply as \fnl hereafter. Slow-roll single-field inflation predicts a nearly vanishing \fnl \citep{Maldacena_2003,Creminelli_2004}. Therefore, detecting a significantly larger value  ($f_{\rm NL} > 1$) would point to a more complex scenario, such as multi-field inflation \citep{Lyth_2003,Byrnes_2010,Pajer_2013}. Currently, the tightest constraints on \fnl come from the bispectrum measured for the anisotropies of the cosmic microwave background,  with $f_{\rm NL} = -0.9 \pm 5.1 \;(68\%\,{\rm c.l.})$ \citep{Planck_2018}. However, this measurement is fundamentally limited by cosmic variance and it is not expected to reach the benchmark of $\sigma(f_{\rm NL}) = 1$,  even with the addition of polarisation data from the CMB \citep{Baumann_2009}. Such a small error would allow us to detect or rule out most multi-field inflation models.

The most promising way to detect local PNG is through galaxy surveys, using the measurement of the so-called scale-dependent bias \citep{Dalal_2008,Slosar_2008,Matarrese_2008}.  This method has been used during the last few years and has resulted in increasingly precise constraints \citep{Slosar_2008,Ross_2012,Giannantonio_2014,Ho_2015, Leistedt_2014,Castorina_2019,Cabass_2022,Rezaie_2023}. The most precise measurement to date from spectroscopic surveys comes from SDSS-IV/eBOSS, which measured $f_{\rm NL} = -12 \pm 21 \, (68\%\,{\rm c.l.})$ \citep{Mueller_2021}. Current spectroscopic galaxy surveys, such as DESI \footnote{\url{https://www.desi.lbl.gov/}} and EUCLID \footnote{\url{https://www.esa.int/Science_Exploration/Space_Science/euclid_overview}}, are expected to reach a level of precision of $\sigma (f_{\rm NL}) \sim 5$ \citep{Laurejis_2011,Giannantonio_2012,Sartoris_2016,DESI_2016}. The next generation of cosmological surveys is expected to reach $\sigma (f_{\rm NL}) < 1$ by combining the information coming from galaxy clustering with other probes, such as HI intensity mapping \citep{LSST_2009,Yamauchi_2014,SKAO_2020, Jolicoeur_2023}. 

Interpreting current and future cosmological surveys requires the use of large simulations. They are routinely used for various purposes, from estimating covariance matrices \citep{Manera_2012,Avila_2018,Zhao_2021} to validating the analysis pipeline for the observational data \citep{Avila_2020, Alam_2021,  Rossi_2020} and studying the theoretical models and observational aspects in a controlled environment \citep[e.g.][]{Ross_2017,Avila_2021, Chan_2022,Riquelme_2022}.  
To date, there has been no adequate simulation to perform these studies in the presence of local PNG for stage IV spectroscopic surveys, given the limited mass resolution of the existing ones \citep{Grossi_2009,Desjacques_2009,Pillepich_2009,Hamaus_2011, Scoccimarro_2012,Wagner_2012}. 

To study the scale-dependent bias with the precision required by current galaxy surveys such as DESI ($\sigma(f_{\rm NL})\sim 5$), the required simulation would need to have $>16\,000^3$ particles in a volume  larger than $(4 \; {h^{-1} {\rm Gpc}})^3$. As this is extremely expensive to run, several techniques have been developed in recent years to increase the statistical significance of simulations. This increase can be obtained either by changing the way the initial conditions are generated \citep[for example the fixed-and-paired technique proposed by][]{Angulo_2016} or by using a set of approximate methods that are computationally less demanding and/or less expensive \citep{Chartier_2021,Ding_2022,Kokron_2022,DeRose_2023}.

First, we needed to clarify whether the fixed-and-paired technique could be used in the presence of local PNG, as fixing the Fourier modes' amplitudes breaks Gaussianity of the initial conditions. However, in \citet{Avila_2022}, we demonstrated that this can also be used  in the presence of local PNG to have a larger effective volume for simulations. In addition, in that work, we introduced a technique further to suppress the noise in the \fnl measurement. This suppression is achieved by matching the phases between two simulations with and without PNG, which removes much of the noise in the measured statistics associated with the cosmic variance. We use the results from \citet{Avila_2022} as a crucial ingredient in the following analysis.

This work presents the \pngsim suite, which includes the largest simulation (in terms of number of particles) to date with non-Gaussian initial conditions, the \textsc{PNG-UNIT}. This simulation evolves $4096^3$ dark matter particles within a cosmological volume of $(1 h^{-1}\, {\rm Gpc})^3$, resulting in a mass resolution of $1.24\times10^{9} \, h^{-1} M_\odot$. Most of the previously developed simulations with local-PNG were run with fewer particles, that is, by over a factor of $18$  \citep[e.g.][]{Nishimichi_2010,Baldauf_2016,Biagetti_2017,Coulton_2023}. Due to this low resolution, it was not possible to use these simulations to study the scale-dependent bias for low-mass halos and galaxies (i.e. $M_{\rm halo} \lesssim 1\times 10^{13} \, h^{-1} \, M_\odot$). After the submission of this paper, \citet{Hadzhiyska_2024} presented a new PNG simulations with the same number of particles as the \textsc{PNG-UNIT}. However, these simulations have a volume of $(2\,h^{-1} {\rm Gpc})^3$, resulting in a mass resolution $\times8$ times lower than the \textsc{PNG-UNIT}. Therefore, their objectives are different and complementary to the project presented in this paper.

The resolution of the \textsc{PNG-UNIT} simulation is enough to finely resolve (with more than $\sim 100$ particles) the halos hosting the emission-line galaxies (ELGs) targeted by DESI and EUCLID $(\sim 10^{11} \, h^{-1} M_\odot)$ \citep{Gonzalez-Perez_2018,Yuan_2023,Rocher_2023,Reyesperaza_2023}. Other DESI tracers (BGS, LRG, QSO) are expected to probe even larger masses, which are also well resolved by this simulation \citep{Jiaxi_2023,Rocher_2023,Prada_2023}. The initial conditions of the \textsc{PNG-UNIT} are generated with amplitudes fixed to their expectation values and phases matched to one of the original \textsc{UNIT} simulations with Gaussian initial conditions \citep{Chuang_2019}. 

We first used the \pngsim suite to constrain the non-Gaussian bias parameter, \bp, for mass-selected halos. This parameter controls how the clustering of halos, or any given tracer of the matter density field (cosmological tracer hereafter), responds to the presence of PNGs. However, \bp is completely degenerate with the amplitude of the non-Gaussianities measured by \fnl \citep{Barreira_2022}. It is, therefore, crucial to have a good understanding of \bp if we aim to measure \fnl from data. Thus, understanding \bp from simulations will be essential. 

An assumption is made for the mass function for halos where the mass only depends on the height of the density peak, namely, $\nu = \delta_c/\sigma(M)$, where $\delta_c$ is the critical overdensity and $\sigma(M)^2$ is the variance of the density field smoothed at mass $M$. On this basis, a theoretical prediction for \bp can be made as a function of the linear bias $b_1$ \citep{Dalal_2008}:
\begin{equation}
    b_\phi = 2\delta_c (b_1 -1).
    \label{eq:universality}
\end{equation}

This result is what is usually called the universality relation. Nevertheless, several works have shown that \bp may not be accurately described by that theoretical expectation \citep{Slosar_2008,Grossi_2009,Hamaus_2011, Scoccimarro_2012,Wagner_2012,Biagetti_2017}. Moreover, it has been demonstrated that \bp may differ depending on the particular cosmological tracer \citep{Barreira_2020, Barreira_2022c}.

In the literature, \bp has been constrained using not only the scale-dependent bias in simulations with PNG but also with other methods, namely the separate universe technique \citep{Biagetti_2017,Barreira_2020, Barreira_2022,Lazeyras_2023}. Although \bp can be measured very precisely with this approach, these simulations cannot validate non-Gaussian analysis tools of, for example,  galaxy clustering, weak lensing statistics and cluster counts given they assume Gaussian initial conditions. Moreover, 
in this paper we find that the errors for the \bp obtained with the separate universe method may be underestimated.

In this work, we divided the halos into 12 pseudo-logarithmically spaced mass bins and studied their clustering by measuring their power spectrum. This statistic is sensitive to the product $b_{\phi} f_{\rm NL}$ through the scale-dependent bias \citep{Dalal_2008,Slosar_2008}. Here, we estimate the variance of the power spectrum for halos in \textsc{PNG-UNIT} from a set of fast mocks generated with \textsc{FastPM} \citep{Feng_2016} with fixed initial conditions. We further improved the precision measuring $b_\phi f_{\rm NL}$ by matching the phases of the initial conditions. In the future, we plan to carry out a study of how \bp depends on the environment and the physics of galaxy formation in a non-Gaussian simulation. 

This paper is structured as follows. A brief review of the theory of galaxy clustering in the presence of local PNG is given in Section \ref{sec:theory}, focusing on the scale-dependent bias.  In Section \ref{sec:simulations}, we present the \pngsim suit of simulations developed for this study. In Section \ref{sec:methods-pk}, we describe the methods employed for defining and relating the different mass bins, estimate the power spectrum variance, and apply the matching technique to improve the constraints on the PNG response parameters. In Section \ref{sec:results}, we present and discuss our main findings, including the constraints on the \bp parameter for mass-selected halos. In Section \ref{sec:sec_convergence}, we study the convergence of our analysis with different sets of simulations and mass resolutions.  The set of tests performed to ensure the robustness of our results are described in Section \ref{sec:robustness_tests}. In Section \ref{sec:priors}, we discuss how our results can be used as a prior for a DESI-like survey and how it would affect the constraints on \fnl. Finally, we present our conclusions in Section \ref{sec:conclusions}.


\section{Theory}\label{sec:theory}

Dark matter halos trace the matter distribution, as these objects are assumed to form from high-density peaks. However, the gravitational collapse of dark matter halos is a non-linear and complex process. On quasi-linear scales, the complexities can be absorbed into a set of operators whose coefficients are called bias parameters \citep{Bardeen_1986,Desjacques_2018}.  In the context of perturbation theory, the density contrast of halos and also other cosmological tracers, such as galaxies, can be written as a function of position and redshift:

\begin{equation}
    \delta_h(\vec{x},z) = \sum_{O} b_{O}(z) O(\vec{x},z)\; .
    \label{eq:bias_expansion}
\end{equation}

The operators, $O$, describe all the fields that may affect the matter density field, while $b_O$ represents the bias parameters associated with each operator.  If only local operators are considered, then the bias expansion can only contain terms with exactly two spatial derivatives of the potential, that is, $O \propto \partial_i\partial_j \phi$ \citep[see][for an extensive review on the bias formalism]{Desjacques_2018}. All the terms proportional to the matter density contrast $\delta_m$ (e.g. $\propto \delta_m^n$ ) are allowed in the local bias expansion given it is related with the potential through the Poisson equation $(\nabla^2\phi\propto\delta_m)$.  The leading order of  \autoref{eq:bias_expansion} can be expressed as:
\begin{equation}   
\label{eq:linear_bias_density}
    \delta_{h} =  b_1  \delta_{m} \; ,
\end{equation}
where $\delta_h$ is the density field of halos, $\delta_m$ is the density field of the matter distribution, and $b_1$ is the linear bias. This expression is only valid on the purely linear regime, which typically is considered those below $k_{\rm max} \lesssim 0.1 \, h\,{\rm Mpc}^{-1}$. By including more terms in \autoref{eq:bias_expansion}, the clustering of the halos can be modelled even in the mildly non-linear regime.  This modelling is unnecessary as the effect of local PNG is mostly encoded in the largest scales (see \S\ref{sec:scale_dependent_bias}). Hence, most of our analysis is focussed on the purely linear regime.

Using the linear bias, $b_1$, from \autoref{eq:linear_bias_density}, the power spectrum of halos can be expressed  in terms of the linear matter power spectrum, $P_{m,m}(k,z)$, as:
\begin{equation}
    P_{h,h}(k,z) = \langle \delta_h \delta_h\rangle = b_1^2(z) P_{m,m}(k,z) \; .
    \label{eq:halo_power_spectrum_linear}
\end{equation}

Below, we describe how PNG may arise from non-standard inflationary models (\S\ref{sec:PNG}) and how it induces a scale dependence in the linear galaxy and halo bias (\S\ref{sec:scale_dependent_bias}).


\subsection{Primordial non-Gaussianities}\label{sec:PNG}

The simplest inflationary models, such as single field slow-roll inflation, predict a nearly Gaussian spectrum of perturbations \citep{Maldacena_2003,Creminelli_2004}. However, many models deviate from these assumptions, which may lead to large primordial non-Gaussianities (PNG) \citep[e.g.][]{Byrnes_2010,Ezquiaga_2022}.  At first order, the amplitude of these deviations from Gaussianity is typically measured in terms of the first-order parameter \fnl.  In the case of Gaussian fields, all the information is contained in the two-point function, and the higher-order statistics, such as the bispectrum and the trispectrum, are determined by the Wick theorem \citep{Desjacques_2018}. However,  this relation is modified in the presence of PNGs, and non-trivial contributions to the higher-order statistics may arise. 

At first order, this \fnl term induces a non-vanishing component in the bispectrum. Specifically, by splitting the primordial potential into a Gaussian and a non-Gaussian part as $\phi = \phi_G + \phi_{NG}$, it can be shown that the leading order contribution to the bispectrum is given by:

\begin{equation}
    \langle\phi_G \phi_G \phi_{NG}\rangle = (2\pi)^3 \delta_D(\vec{k}_1+\vec{k}_2+\vec{k}_3) \mathcal{B}_{\phi}(\vec{k}_1, \vec{k}_2,\vec{k}_3) 
    \label{eq:leading_bispectrum}
\end{equation}
where $\delta_D$ is the Dirac delta in 3D and $ \mathcal{B}_{\phi}(\vec{k}_1, \vec{k}_2,\vec{k}_3)$ is the primordial bispectrum. Depending on the shape induced in the primordial bispectrum, three types of PNGs have been traditionally considered: local, equilateral and orthogonal. Nevertheless, the landscape of possible non-Gaussianities is much richer and more complex. 

This work focuses only on the local type, one of the simplest forms of PNGs. It can be described by the following parameterisation of the primordial gravitational potential \citep{Salopek_1990, Komatsu_2001}:
\begin{equation}
    \phi(\vec{x}) = \phi_G(\vec{x}) + f_{\rm NL} \left(\phi_G(\vec{x})^2 - \langle \phi_G(\vec{x})^2 \rangle\right) \; , 
    \label{eq:potential_local_png}
\end{equation}
where $\phi_{G}$ is the Gaussian primordial gravitational potential. The total primordial potential, $\phi$, is modified at quadratic order by a term proportional to $f_{\rm NL}\phi_G^2$ and now is non-Gaussian. 

One finds that a non-vanishing component to the three-point function is induced from the expression for the local PNGs given in  \autoref{eq:potential_local_png}. In this case, the bispectrum peaks in the squeezed limit ($k_3 \ll k_1,  k_2$) and is given by
\begin{equation}
     \mathcal{B}_{\phi}(\vec{k}_1, \vec{k}_2,\vec{k}_3) = 2 f_{\rm NL} P_\phi(k_1) P_\phi(k_2) + (2 \;{\rm cyc.\; perm}),
    \label{eq:bispectrum_squeezed}
\end{equation}
where $P_\phi(k)$ is the power spectrum of the primordial  potential $\phi$,  and $(2 \;{\rm cyc.\; perm})$ are two cyclic permutations of the term $P_\phi(k_1) P_\phi(k_2)$.

The perturbations in the potential lead to a density contrast in the matter field through the Poisson equation. Then, the non-Gaussianities in the primordial potential are induced also in the matter density field.  These fields can be related in the Fourier space via:
\begin{equation}
    \delta(k,z) = \alpha(k,z) \, \phi(k,z)\; ,
    \label{eq:delta_2_phi}
\end{equation}
where $\alpha(k,z)$ is defined as
\begin{equation}
\alpha(k,z) = \frac{2 D(z)}{3 \Omega_{\rm m,\; 0}} \frac{c^2}{H_0^2} \frac{g(0)}{g(z_{\rm rad})}k^2 T(k) \; .
\label{eq:alpha_k}
\end{equation}
The parameter $\Omega_{\rm m,\; 0}$ is the matter density parameter today, $c$ is the speed of light, and $H_0$ is the Hubble parameter today. The transfer function $T(k)$ is normalised such that $T(k\rightarrow0)=1$. The growth factor $D(z)$ is also normalised to $D(z=0)=1$. The term $\frac{g(z_{\rm rad})}{g(0)}$ (with $g(z)=(1+z)D(z)$) arises from the normalisation we assume for $D(z)$ \citep{Mueller_2018}. If the growth factor is normalised to $(1+z)^{-1}$ during matter domination the term $\frac{g(z_{\rm rad})}{g(0)}$ can be omitted. For the cosmology described in \citet{Planck_2015} and we assume throughout the paper, we find that $\frac{g(z_{\rm rad})}{g(0)} = 1.275$.


\subsection{Scale-dependent bias}\label{sec:scale_dependent_bias}

In the presence of local PNG, the leading-order term proportional to $f_{\rm NL} \phi$ in the bias expansion (\autoref{eq:bias_expansion}) gives 

\begin{equation}
\centering
    \delta_h (k,z) = b_1(z) \delta_m(k,z) + b_\phi f_{\rm NL}\phi = \left[b_1(z) + \frac{f_{\rm NL} b_\phi(z)}{\alpha(k,z)} \right] \delta_m(k,z).
    \label{eq:bias_relation_bphi}
\end{equation}
Here $b_1$ is the linear bias, and $\alpha(k,z)$ is given in \autoref{eq:alpha_k}. \bp is the bias parameter associated with the $\phi$ operator. This term only appears in the presence of PNG, as it is associated with the primordial bispectrum (\autoref{eq:bispectrum_squeezed}), which vanishes in the case of Gaussian initial conditions. The term $\alpha$ introduces the $1/k^2$ dependence of $\delta_h$ on the scale $k$.

Using this new bias expansion, the power spectrum for dark matter halos, or any other biased tracer of the total matter density,  can be computed similarly to \autoref{eq:halo_power_spectrum_linear}. The resulting halo power spectrum is given by
\begin{equation}
    P_{h,h}(k,z) = \left[b_1(z) + \frac{f_{\rm NL} b_\phi(z)}{\alpha(k,z)} \right]^2 P_{m,m}(k,z).
    \label{eq:power_spectrum_scale_dependent_bias}
\end{equation}
The relationship between $P_{h,h}(k,z)$ and $P_{m,m}(k,z)$ now depends on the scale due to the term $\alpha(k,z)$, where the leading-order effect is proportional to $1/k^2$. This proportionality is what is commonly called the `scale-dependent bias'.

\begin{figure}
\includegraphics[width=\columnwidth]{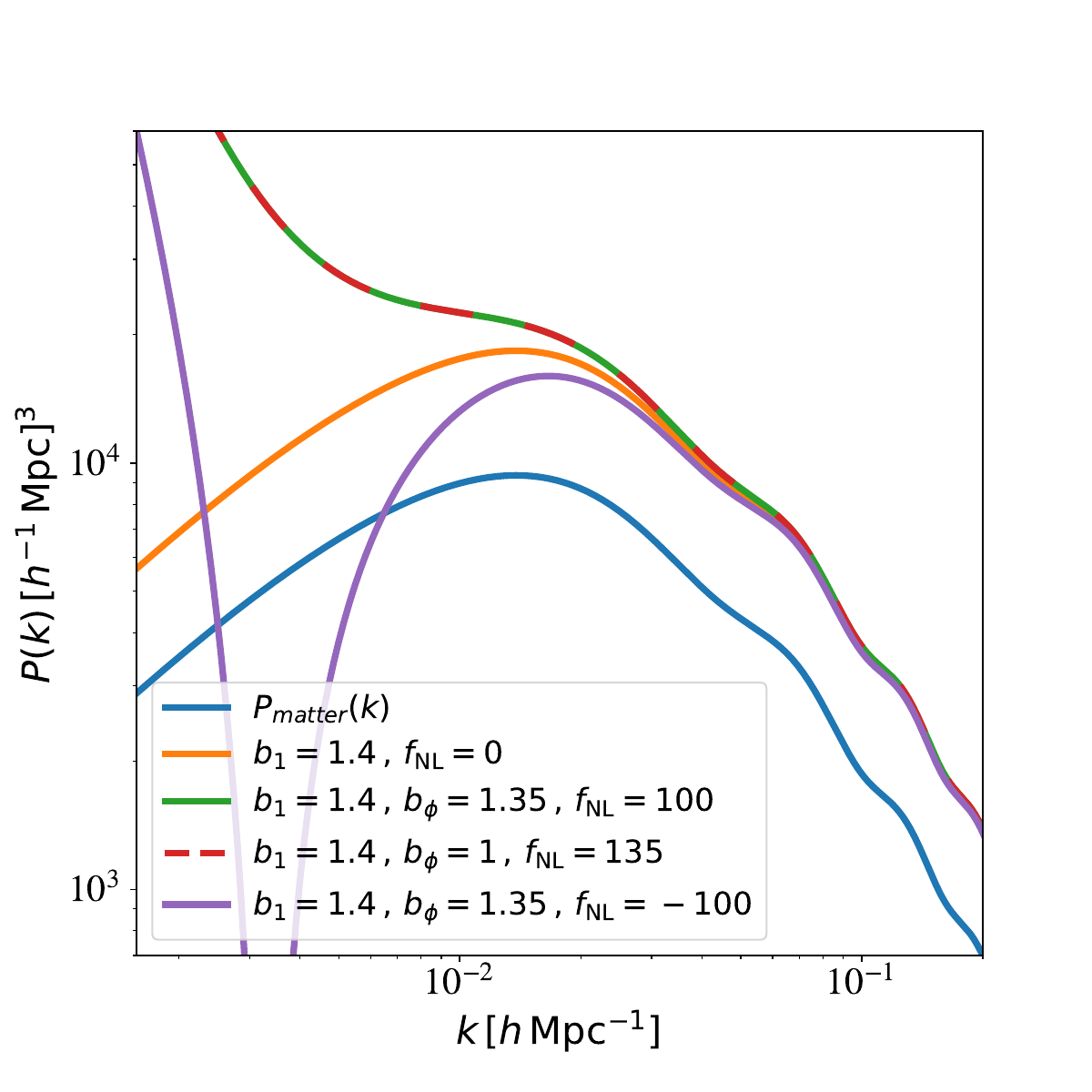}
 \caption{Theoretical power spectrum obtained from \autoref{eq:power_spectrum_scale_dependent_bias}. The blue line shows the linear matter-matter power spectrum for the \textsc{UNIT} cosmology at $z=1$. The $b_1$ parameter is the same for all cases, and we chose it to be similar to the one expected for the DESI emission-line galaxies (ELGs) at that redshift \citep{DESI_2016} . The orange line corresponds to the case of a biased tracer with Gaussian initial conditions. The green line shows the effect of a non-Gaussian contribution with $f_{\rm NL} = 100$ and a PNG-response parameter \bp given by \autoref{eq:bphi_p} with $p=1$. The red line displays how a different value of \fnl can lead to the same signal if the \bp parameter does not follow the universality relation. Red and green lines are on top of each other since the product $b_\phi f_{\rm NL}$ is the same. Finally, the purple line shows the kind of signal we expect in case $b_\phi f_{\rm NL } < 0$.}
 \label{fig:1_model_scale_dependent_bias}
\end{figure}

However, different values of \fnl may produce the same signal depending on the value of the bias parameter \bp as they are perfectly degenerate. This issue has already been discussed in the literature \citep{Barreira_2022}.

The parameter \bp, by definition, represents the response of the abundance of halos to the presence of perturbations in the primordial potential $\phi$. As local PNG induces a mixing between the long-wavelength and short-wavelength perturbations (\autoref{eq:potential_local_png}), in the context of peak-background split theory \citep{Kaiser_1984,Bardeen_1986} it can be demonstrated that \bp can be thought of also as the response of the number density of halos to a rescaling of the amplitude of the primordial power spectrum \citep{Desjacques_2018,Barreira_2022b}:
\begin{equation}
    b_\phi  =\frac{1}{\bar{n}_h}\frac{\partial \bar{n}_h}{\partial \log (f_{\rm NL } \phi) } = \frac{1}{\bar{n}_h}\frac{\partial \bar{n}_h}{\partial \log \mathcal{A}_s } \;,
\label{eq:bphi_partial_As}
\end{equation} 
where $\bar{n}_h$ is the mean number of halos and $\mathcal{A}_s$ is the amplitude of the primordial scalar perturbations. Then, if a universal mass function is assumed, one can derive a theoretical prediction for the value of \bp \citep{Dalal_2008,Slosar_2008,Desjacques_2018}:

\begin{equation}
    b_{\phi} = 2 \delta_c \left(b_1 - p\right) \; .
\label{eq:bphi_p}
\end{equation}
Here, $\delta_c = 1.686$ is the density threshold for the spherical collapse, and $p$ may take different values depending on the population of objects under consideration. By assuming a universal mass function, then $p=1$ for all tracers of the matter distribution \citep{Dalal_2008,Slosar_2008}. This expression with $p=1$ is usually called the universality relation for \bp (noted as the universality relation hereafter). 

Recent studies have shown that this relation does not always accurately describe the scale-dependent bias. For instance, \citet{Slosar_2008} argued that for halos that have recently undergone a major merger process, \bp is better described by $2\delta_c(b_1 -1.6)$, while \citet{Barreira_2020} showed that for galaxies selected by their stellar mass, the relation with $b_1$ is better described by $2\delta_c (b_1 -0.5)$.  Moreover, \bp may depend not only on the halo mass (through $b_1$) but also on other properties such as the halo concentration \citep{Lazeyras_2023,Fondi_2023}. 

In \autoref{fig:1_model_scale_dependent_bias}, we present the expected power spectrum of dark matter halos at linear order in perturbation theory for different values of the non-Gaussian parameter, \fnl. We chose the value of $b_1$ to be similar to the one expected for the emission-line galaxies (ELGs), a tracer sample targeted by the current galaxy surveys \citep{Jiaxi_2023}. For the Gaussian case, the halo power spectrum is just a shifted version of the matter-matter power spectrum in logarithmic scale, with the shift determined by the linear bias parameter $b_1$, as given by \autoref{eq:halo_power_spectrum_linear}. At large scales, the term proportional to $b_\phi f_{\rm NL}$, which appears only for the case with local PNG, induces an enhancement for $b_\phi f_{\rm NL} >0$ (red and green lines), and suppression if $b_\phi f_{\rm NL} <0 $ (purple line) of the power spectrum compared with the Gaussian case. When going to even larger scales, the term proportional to $(b_\phi f_{\rm NL})^2$ dominates, and an enhancement of the power spectrum is shown for both cases, $b_\phi f_{\rm NL} >0 $  and $b_\phi f_{\rm NL} <0 $.

The scale-dependent bias is only sensitive to the product $b_\phi f_{\rm NL}$. Because of this, $f_{\rm NL}$ cannot be extracted from this unless we have prior knowledge of \bp \citep{Barreira_2022}, as illustrated in \autoref{fig:1_model_scale_dependent_bias}, where the green and red lines show the same signal although the \fnl value is different.


\section{Simulations}\label{sec:simulations}
The \textsc{UNITsims}\footnote{\url{http://www.unitsims.org/}} simulation suite \citep{Chuang_2019} aims to model the large-scale structure of the Universe to validate  theoretical models and analysis pipelines for extracting information about cosmological parameters in the current galaxy surveys. For this purpose, simulations need an effective volume equivalent to or larger than these surveys and a mass resolution high enough to resolve the dark matter halos hosting the targeted galaxies and tracers.

\begin{table}
\centering
\caption{Cosmological parameters of all \textsc{UNITsims} and \textsc{PNG-UNITsims} simulation suite \citep{Planck_2015}.}
\begin{tabular}{lr}
\hline
$\Omega_{\rm m}$             &    0.3089 \\
$\Omega_{\rm b}$             & 0.0486 \\
$\Omega_\Lambda$             &    0.6911 \\
$h \equiv H_0$/(100 km\,s$^{-1}$\,Mpc$^{-1})$ &  0.6774\\
$\sigma_8$                   &    0.8147 \\
$n_{\rm s}$                  &    0.9667 \\
$f_{\rm NL}$                 &    0, 100 \\
\hline
\end{tabular}
\label{tab:cosmology}
\end{table}

\begin{table*}
\centering
%
\caption{Characteristics of the main simulations used throughout the present paper.}
\begin{tabular}{@{}cc|ccc@{}}
\cmidrule(l){2-5}
                                        & \textsc{UNIT}                         & \textsc{PNG-UNIT}                 & \textsc{LR-UNIT}            & \textsc{FastPM} mocks      \\ \midrule
Code                                    & L-Gadget 2                            & L-Gadget 2                        &  L-Gadget 2                 & \textsc{FastPM}             \\
$N_{\rm particles}$                     & $4096^3$                              & $4096^3$                          & $2048^3$                    &  $2048^3$                   \\
$m_{p}\; (h^{-1}\,M_\odot )$            & $1.24\times10^{9}$                    & $1.24\times10^{9}$                & $9.97\times10^{9}$          &  $9.97\times10^{9}$         \\
$f_{\rm NL}$                            & 0                                     & 100                               & 0 and 100                   &  0 and 100                \\
Halo finder                             & Rockstar                              & Rockstar                          & Rockstar                    &  FoF                      \\ 
Softening length $(h^{-1} {\rm kpc})$   & 6.1                                   & 6.1                               & 12.2                        &  244.14\hyperlink{note}{*}                       \\ \bottomrule
\end{tabular}
\tablefoot{We refer to the simulations as \textsc{PNG-UNIT}/\textsc{UNIT} while we leave the names \textsc{PNG-UNITsims}/\textsc{UNITsims} for the simulation suites. From top to bottom: the gravity solver, the number of dark matter particles, the particle mass, the values of \fnl set in the initial conditions, the halo finder and the softening length. All the simulations have a box size of $L_{\rm box} = 1 \, {h^{-1} \,{\rm Gpc}}$. The cosmological parameters and the input power spectrum are the same (except for \fnl) between all the simulations. \hypertarget{note}{*} \textsc{FastPM} is a particle-mesh code, so the cell size gives the force resolution, and this is the number we report in the table as softening length for \textsc{FastPM}.}
\label{tab:sim_characteristics}
\end{table*}

Most state-of-the-art simulations, including the \textsc{UNIT} simulation (\S\ref{sec:unitsim}), assume Gaussian cosmology and  cannot be used directly to study the effect that PNG has on the distribution of large-scale structure at later cosmic times. To address this, we have developed the \textsc{PNG-UNIT} (Section\ref{sec:png-unitsim}), together with two lower-resolution full N-body simulations (\textsc{LR-UNIT}, \S\ref{sec:lr-unitsim}) and a set of $100$ fast N-body \textsc{FastPM} mocks (\S\ref{sec:fastpm-mocks}). We developed another set of smaller simulations to perform further robustness tests on our analysis, which we discuss in section \ref{sec:supporting_simulations}.

All simulations presented in this paper share the  same cosmology \citep{Planck_2015}, as we summarise in  \autoref{tab:cosmology}. We outline the characteristics of the simulations in \autoref{tab:sim_characteristics}. Moreover, we focus our analysis on the snapshots at redshift $z=1.032$.

Throughout the paper, we refer to the suite as \textsc{PNG-UNITsims}/\textsc{UNITsims} and to the $4096^3$ particle simulation as \textsc{PNG-UNIT}/\textsc{UNIT}.


\subsection{The \textsc{UNITsims}} \label{sec:unitsim}
The \textsc{UNITsims} are a set of two pairs of state-of-the-art, dark matter-only, full N-body simulations with fixed-and-paired initial conditions \citep{Angulo_2016,Chuang_2019}. Each of them tracks the evolution of $4096^3$ particles in a periodic box with a side  length of $L=1\,h^{-1} \, {\rm Gpc}$ from redshift $z=99$ down to $z=0$.  The  mass resolution  is  $m_p = 1.24\times10^{9}\; h^{-1}\,M_\odot$, so the least massive halos expected to host galaxies targeted by current spectroscopic surveys, such as EUCLID and DESI,  are well resolved in these simulations \citep{Cochrane_2017,Gonzalez-Perez_2018,Chuang_2019, Knebe_2022}. In addition to the new simulations we run, we are also using in this work one of the original \textsc{UNITsims}, for reference, the one labelled FixAmpInvPhase\_001.

The initial conditions were set using the second-order Lagrangian Perturbation Theory (2LPT), implemented in \textsc{FastPM} \footnote{\url{https://github.com/fastpm/fastpm}} \citep{Feng_2016}. The required initial power spectrum was computed using the Boltzmann solver \textsc{CAMB}\footnote{\url{https://camb.info/}} \citep{Lewis_2000}.

The dark matter particles were evolved using the N-body code \textsc{L-Gadget2} \citep{Springel_2005}. This code uses a Tree-PM algorithm to compute the forces between particles by splitting the Newtonian potential into long-range and short-range components.  We employed the non-public version of the code \textsc{L-Gadget-2} as it is highly optimised regarding memory usage. This code has been widely used for generating many large-volume and high-resolution simulations such as the \textsc{Millennium} run \citep{Springel_2005b} and the \textsc{Multidark} suite \citep{Klypin_2016}. The Plummer softening length was set to $\epsilon = 6 \, {\rm kpc}/h$ and then the particles were evolved from $z=99$ down to $z=0$, stored in a total of $129$ snapshots logarithmically spaced in the scale factor. 

Dark matter halos  were identified using the publicly available code \textsc{Rockstar}\footnote{\url{https://bitbucket.org/gfcstanford/rockstar/src/main/}} \citep{Behroozi_2012a}. This code uses the particles' 6D phase space information to identify the dark matter halos. These structures are identified as having at least 20 particles per halo.  We used \textsc{ConsistentTrees} to track their merger histories and build the merger trees.  More details regarding these simulations can be found in \citet{Chuang_2019}.


\subsection{The \textsc{PNG-UNITsims} }\label{sec:png-unitsim}
We developed the new \pngsim suite for this paper with the same features as the original \textsc{UNITsim} simulations but with non-Gaussian initial conditions. Specifically we set $f_{\rm NL} = 100$. The suite consists of a high-resolution simulation (\S\ref{sec:png-unitsim-simulation}), a lower resolution version of it (Section\ref{sec:lr-unitsim}) and a set of $100$ fast mocks generated with \textsc{FastPM}, half of them with \fnl$=0$ and the other half with \fnl$=100$ (\S\ref{sec:fastpm-mocks}). 

To generate the initial conditions of all of these simulations, we have used the \textsc{FastPM} code. In the first place, the initial conditions are generated for $\phi$ in the same way as if the simulation were Gaussian. Indeed, at this point, the amplitudes of the Fourier modes can be fixed to their expectation value as in \citet{Angulo_2016}. Then, to induce the local PNGs, the primordial potential is computed from the density field using \autoref{eq:delta_2_phi} and then the transformation given in \autoref{eq:potential_local_png} is applied. This procedure is detailed and validated in \citet{Avila_2022}.


\subsubsection{The \textsc{PNG-UNIT}} \label{sec:png-unitsim-simulation}
The \textsc{PNG-UNIT} simulation is the main product of this work as it is the largest full N-body simulation in terms of number of particles with primordial non-Gaussianites to date. It is a dark matter-only, full N-body simulation with local primordial non-Gaussianities. As the original  \textsc{UNIT} simulation, it tracks the evolution of $4096^3$ particles in a periodic box with a side  length of $L=1\,h^{-1} \, {\rm Gpc}$ from redshift $z=99$ down to $z=0$. The details are shown in \autoref{tab:sim_characteristics}. Given the large volume and the high mass resolution, this simulation is a unique laboratory to validate models of galaxy clustering in the presence of local PNG.

The value of \fnl used for generating the initial conditions is set to $f_{\rm NL} = 100$. This large value of \fnl is already ruled out by Planck at  $\sim 20 \sigma$ \citep{Planck_2018}.  However, we wanted a strong signal from the scale-dependent bias to understand this effect better. With a more realistic \fnl, given our limitation in scales with minimum wavenumber $k_{\rm min} = k_{f} = 0.006\,  h \, {\rm Mpc}^{-1}$ (where $k_f$ is the fundamental wavenumber), we would not be able to detect the PNG signal in the halo power spectrum.

In addition, the initial conditions of the \textsc{PNG-UNIT} were generated using the fixed-and-matched technique described in \citet{Avila_2022}. The amplitudes were set to the expectation value as in \citet{Angulo_2016}. This method can be safely applied also in simulations with local primordial non-Gaussianities as demonstrated in \citet{Avila_2022}. Then, the phases were chosen to be the same as in the reference  Gaussian \textsc{UNIT} simulation. The N-body code and the halo finder were run using the same configuration as  for the \textsc{UNITsims}. We focus our analysis on the snapshot at redshift $z=1$.


\subsubsection{The \textsc{LR-UNIT} simulations}\label{sec:lr-unitsim}
In addition to the high-resolution simulation presented above, for the purposes of this paper, we have developed two new lower-resolution versions of the previous ones (with $f_{\rm NL} = 0$ and $f_{\rm NL} = 100$). The primary purpose of these \textsc{LR-UNIT} simulations is to understand the effects of mass resolution better (see \S\ref{sec:convergence_lr}). For this reason, we use the same configuration as the high-resolution simulations. The only difference is the resolution (see \autoref{tab:sim_characteristics}).

As we used the same code and seed to generate the initial conditions, all the long-wavelength modes were set to the same values as in the high-resolution simulations. Thus, the same region of the Universe is modelled in these low-resolution simulations, and the same structures can be identified in both sets of simulations. 


\subsection{\textsc{FastPM} mocks}\label{sec:fastpm-mocks}
We have also generated a set of $100$ fast mocks using \textsc{FastPM} \citep{Feng_2016}. These mocks were split into two sets of $50$: one set had Gaussian initial conditions ($f_{\rm NL} = 0$) while the other set had non-Gaussian initial conditions ($f_{\rm NL} = 100$).

\textsc{FastPM} is a code that combines perturbation theory with a particle--mesh (PM) approach to quickly generate cosmological simulations at the cost of poor resolution of the short-wavelength modes. This code is particularly useful for generating many simulations quickly, which can be used, for example,  for estimating covariance matrices at a low computational cost. In this work, we use \textsc{FastPM} mocks to estimate the variance of the power spectrum of the halos in simulations with fixed initial conditions as well as for obtaining the correlation coefficients required by the matching technique as we discuss in Section \ref{sec:matching}. 

Each simulation box contains $2048^3$ dark matter particles within a cosmological volume of $1\;(h^{-1}\,{\rm Gpc})^3$, the same volume as the \textsc{UNIT} simulation. The mass resolution is $m_p = 9.97\times10^{10} \; h^{-1}\,{\rm M_\odot}$, which is $8$ times coarser than the high-resolution simulations (same resolution as \textsc{LR-UNIT} simulations) but still sufficient for this work.  The force resolution in the PM approach is given by the grid size used to compute the potential and the forces. We use a regular grid of $4096^3$ cells, giving a spatial resolution of $244.14 \; h^{-1} {\rm kpc}$.

Following the same procedure as the full N-body simulations, the initial conditions were set at $z=99$ using second-order Lagrangian perturbation theory with fixed amplitudes. The particles were then evolved to $z=1$ using $100$ linearly spaced timesteps.

We use the friends-of-friends (FoF) algorithm to generate the halo catalogues, which is implemented by default in \textsc{FastPM}. The linking length is set to the standard value of $l_l = 0.2 \, l_p$, where $l_p$ is the mean interparticle distance.

The difference between the halo finder used for the \textsc{FastPM} and the \textsc{PNG-UNIT} catalogues leads to a different mass definition for dark matter halos. This issue is addressed in Section \ref{sec:mass_bins}.


\subsection{Supporting simulations}\label{sec:supporting_simulations}

In addition to the \textsc{PNG-UNIT} and the \textsc{FastPM} mocks presented above, we developed an additional set of simulations mainly aimed at checking the robustness of our results. We want to check if the results obtained from fitting \bp in \autoref{eq:power_spectrum_scale_dependent_bias} are consistent with measuring \bp using the separate universe technique (see \S\ref{sec:convergence_sep}). Hence, the set of simulations is split into two groups: high-resolution (HR) and low-resolution (LR) separate universe simulations.

Each group consists of 3 simulations with $512^3$ dark matter particles within a periodic cubic box of size $L_{\rm box, \; HR} = 67.5 \, h^{-1} {\rm Mpc}$ and $L_{\rm box, \; LR} = 250 \, h^{-1} {\rm Mpc}$ for the high-resolution and low-resolution groups respectively. This configuration leads to a mass resolution which is eight times better and eight times worse, respectively, compared to the baseline \textsc{UNITsims}. Again, the fiducial cosmology is the same as the one of the \textsc{UNIT} simulations, summarised in \autoref{tab:cosmology}. Following a similar procedure as shown in \citet{Barreira_2022}, we produced three simulations with the same initial conditions but varied the $\mathcal{A}_s$ (or equivalently $\sigma_8$) by $5\%$. This variation allows us to numerically measure \bp by computing the derivative (\autoref{eq:bphi_sep_uni}), as we discuss in Section \ref{sec:convergence_sep}. 

For all these simulations, we evolved the particles using the \textsc{L-Gadget 2} code, as in the case of the high-resolution simulations and we set the Plummer softening length to $1/40$ times the mean interparticle distance, which varies depending on the mass resolution of the simulations.

\subsection{Compared halo mass functions}\label{sec:hmf}
\begin{figure}
 \includegraphics[width=\columnwidth]{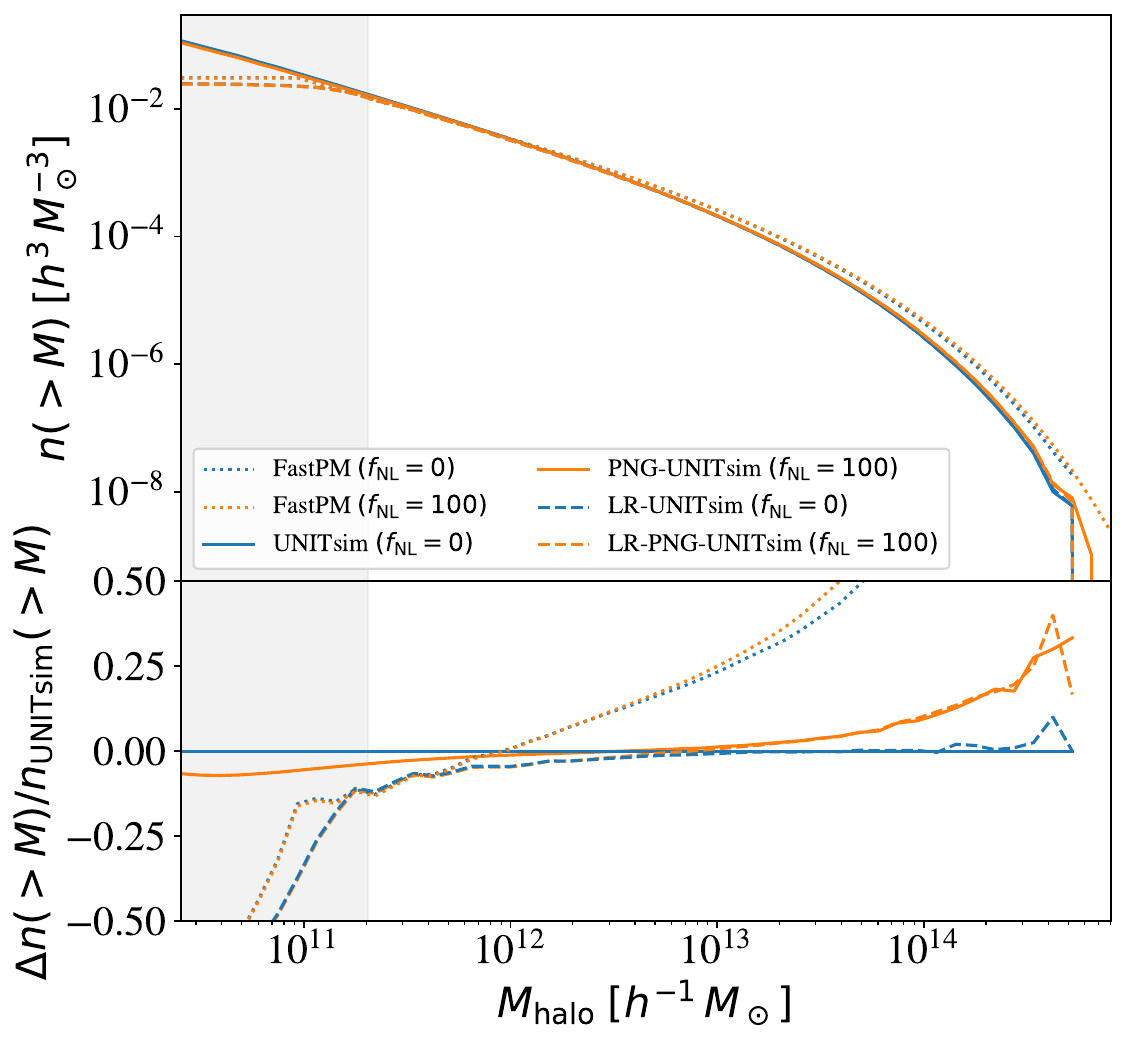}
 \caption{Halo mass functions for the simulations described in Section \ref{sec:simulations} (top). Ratio of halo mass functions relative to the Gaussian \textsc{UNIT} simulation as the reference (bottom). Shaded areas represent the  $M = 20 m_{\rm part}^{\rm LR}$ mass threshold for the low-resolution simulation. The minimum mass shown here represents the 20 particle limit for the higher resolution simulations.}
 \label{fig:2_halo_mass_function}
\end{figure}

\autoref{fig:2_halo_mass_function} shows the cumulative mass function of the halos from the \textsc{PNG-UNIT} and \textsc{LR-UNIT} simulations. The mass function has been calculated from $2.48\times 10^{10}\, h ^{-1} M_\odot$ ($20$ times the particle mass in \textsc{PNG-UNIT}) up to $1\times\, 10^{15} h ^{-1} M_\odot$ in 50 logarithmically spaced bins. We find that the low-resolution simulations converge to the high-resolution ones at $<5\%$ for $M \sim  100\, m_{\rm part}^{\rm LR}$, with $m_{\rm part}^{\rm LR}$ being the particle mass of the lower resolution simulation. The shaded area represents the threshold of $20$ particles per halo, below which the mass of the halos is unreliable in the low-resolution simulations.

The first thing we remark is that the mass function of the \textsc{FastPM} mocks differs significantly from the ones from the full N-body simulations. The reason is that the mass definition is different since for the \textsc{FastPM} mocks, halos were identified with a FoF algorithm while we used \textsc{Rockstar} for the halos in the other simulations. This difference highlights the need for establishing a correspondence between the full N-body halos and the \textsc{FastPM} halos. We devote Section \ref{sec:mass_bins} to this task. 

We checked that the differences in the mass function are mainly due to the different halo-finding algorithms by running Rockstar on one of \textsc{FastPM} mocks. Running both halo finders on the same \textsc{FastPM} mock, we find that for high masses, the mass function matches that of the full N-body simulations, while we find a deficit in the number of low-mass halos; this is probably due to differences in the resolution of the forces between the two approaches.

Moreover, we find an increment of the high-mass halos in the $f_{\rm NL} = 100$ simulations with respect to the Gaussian ones. This increment was expected due to the change of the probability density function (PDF) of $\delta$ due to the presence of local PNG \citep{Matarrese_2000, LoVerde_2008}. 

The Gaussian \textsc{UNIT} simulation has an excess of low mass halos compared with the other simulations due to an issue with the initial conditions affecting the older version of the \textsc{FastPM} code used to produce that simulation. This problem produced an excess power in the power spectrum of the initial conditions at scales between $k_{\rm nyquist}/4$ and $k_{\rm nyquist}/2$ of the order of $10\%$. This results in an overproduction of halos of the same order of $\sim 10\%$ and explains the differences between the \textsc{UNIT} and the \textsc{PNG-UNIT} simulations at the mass function level. However, this does not significantly affect the halo clustering measurements compared to the low-resolution simulations, as we explain in Section \ref{sec:convergence_lr}.

\section{Uncertainty of the Power Spectrum in simulations with fixed-and-matched initial conditions} \label{sec:methods-pk}

In this work, we study the PNG  parameters \bp and \fnl as a function of halo mass. We measure them using the scale-dependent bias induced by PNG in the halo power spectrum (Section\ref{sec:scale_dependent_bias}). The dark matter auto-power spectra of the simulations are obtained by interpolating the particles onto a regular grid with $N=512^3$ cells using a cloud-in-cell (CIC) scheme \citep{Hockney_1981} with a $k$-bin width equivalent to the fundamental wavenumber of the box $(k_f = 2\pi/L_{\rm box} = 0.00628\, h\,{\rm Mpc}^{-1})$. We have used the \textsc{Python} library \textsc{Nbodykit}\footnote{\url{https://nbodykit.readthedocs.io/en/latest/}} \citep{Hand_2018} for this task.   The halo power spectrum is computed similarly but using the halo positions instead of the dark matter particles. In addition, we use a large set of fast mocks to estimate the variance in the power spectrum from the full N-body simulations. This approach allows us to consider the variance suppression due to the initial conditions (see \S\ref{sec:variance_estimation}).  However, fundamental differences exist between the fast mocks and the \textsc{PNG-UNIT}. The main differences are how the halos are defined and the mass resolution of the simulations. We now discuss how to connect the halos (and, hence, the measured power spectrum) from the full N-body simulation with the corresponding ones from the fast mocks. 


\subsection{Adapting \textsc{FastPM} mass bins to \textsc{UNITsim}}\label{sec:mass_bins}

Throughout this work, we only consider main halos, which are not substructures of a larger halo. For simulations where we run \textsc{Rockstar} (see \autoref{tab:sim_characteristics}), the halo mass is estimated from a spherical overdensity up to the threshold of $\Delta = 200 \rho_c$, where $\rho_c$ is the critical density \citep{Behroozi_2012a}. For the \textsc{FastPM} mocks, the halo mass is the sum of the particle mass of its FoF components. This mass definition leads to some differences when choosing the mass bins. The edges of the mass bins are chosen with pseudo-logarithmic spacing for \textsc{UNIT}, as summarised in \autoref{tab:mass_bins}. 

The mass definitions in the full N-body simulations differ significantly from those for the fast mocks. To relate halos between these two approaches, we aim to reproduce simultaneously the clustering and the abundances of the \textsc{UNIT} halos with the \textsc{FastPM} mocks. To this end, for each mass bin of the \textsc{UNITsims}, we  consider a maximum mass for the halos in the \textsc{FastPM} mocks and select the following $N$ most massive halos to recover the \textsc{UNIT} number density. Using a least-squares algorithm, we find the optimal maximum mass for \textsc{FastPM} that reproduces the \textsc{UNIT} power spectrum for $k<0.1\, h\, {\rm Mpc}^{-1}$. We take the minimum and the maximum mass from the resulting bins and apply these cuts for the $f_{\rm NL} = 100$ simulations. 

\begin{figure}
 \includegraphics[width=\columnwidth]{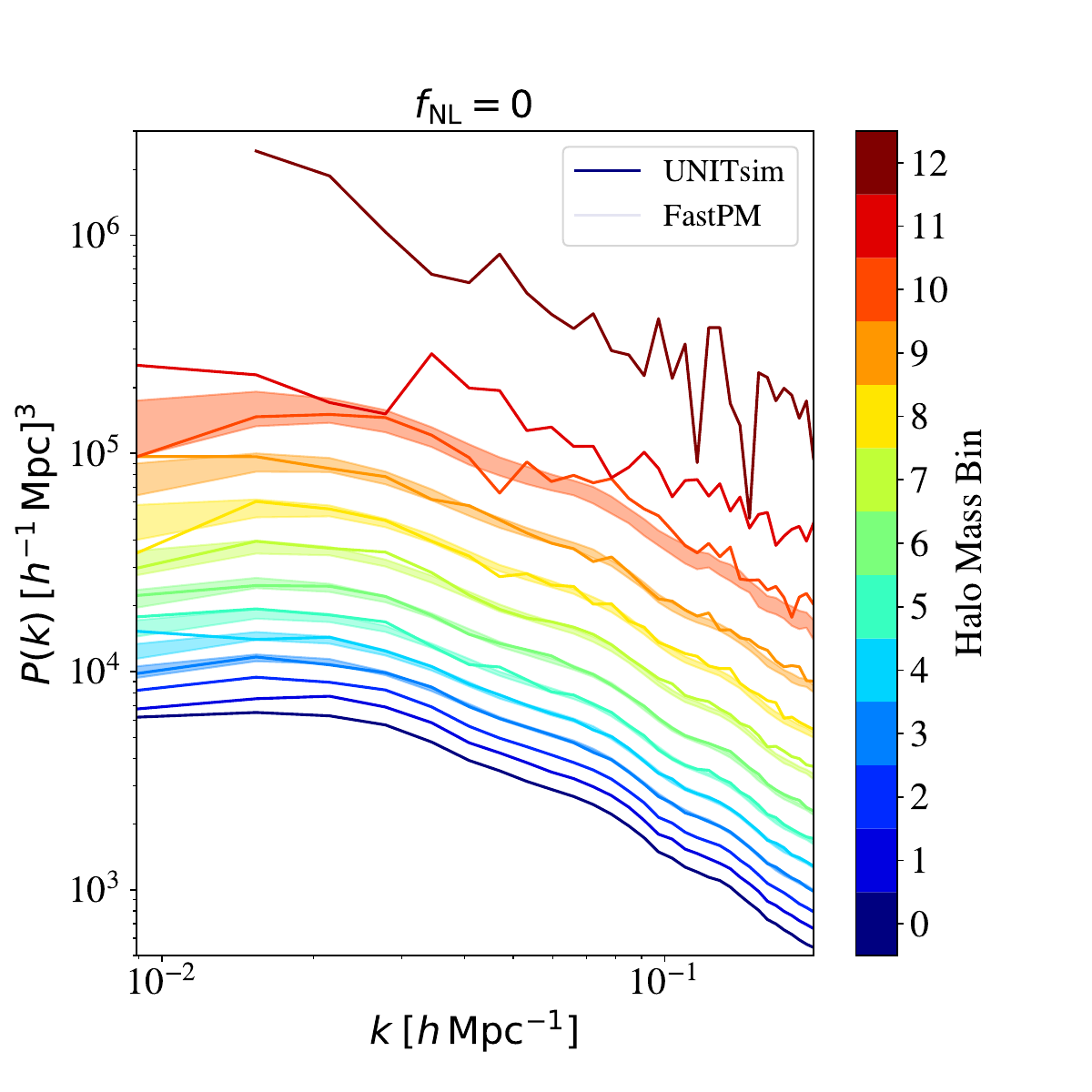}
 \includegraphics[width=\columnwidth]{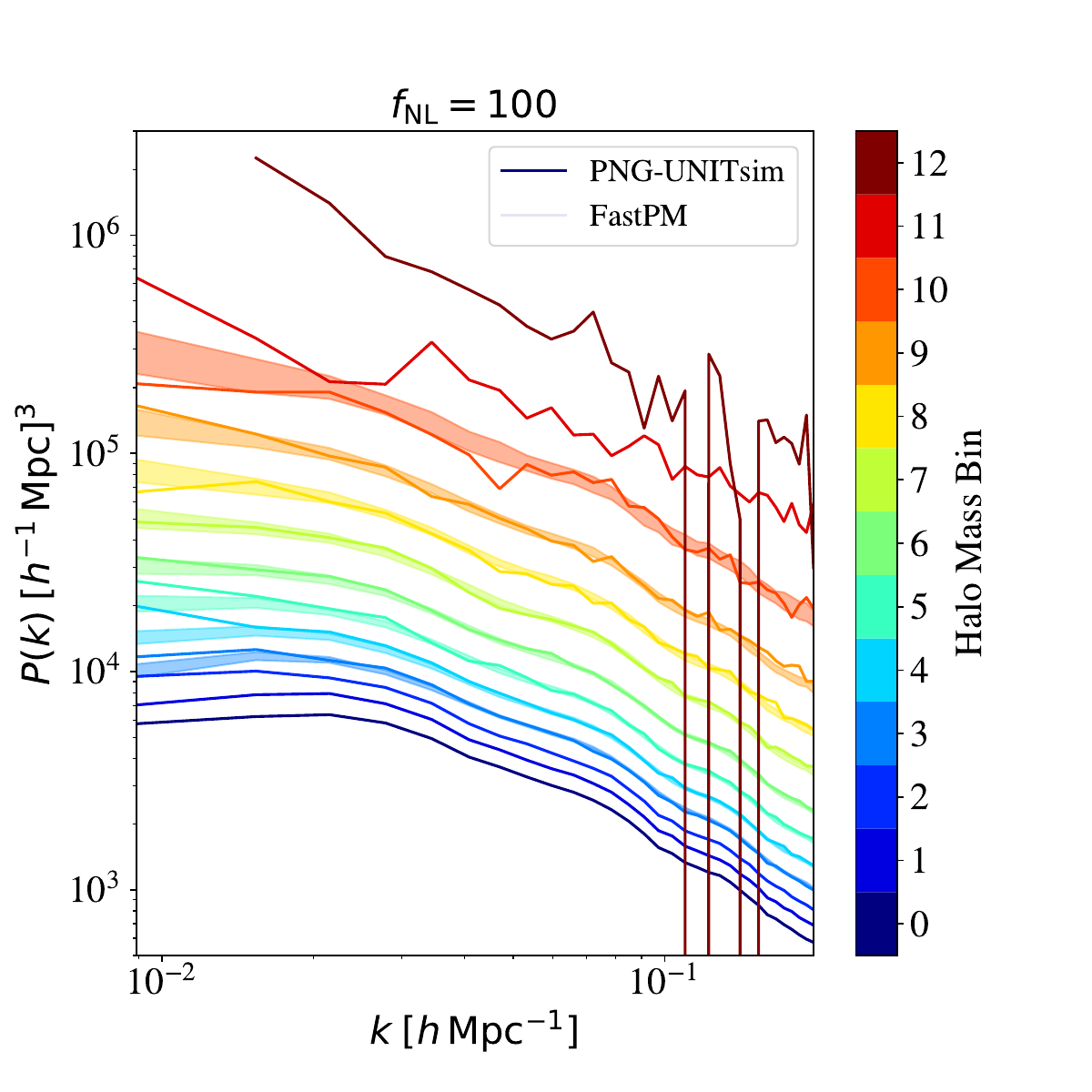}
 \caption{Solid lines represent the power spectrum of \textsc{PNG-UNIT} halos within the different halo mass bins (see \autoref{tab:mass_bins} for the mass ranges). Shaded regions show the standard deviation around the mean power spectrum of the \textsc{FastPM} mocks, generated using halos in mass bins that reproduce the clustering and the abundance of the N-body simulation (see Section\ref{sec:mass_bins}). There are no halos in the two lowest-mass bins (1 and 2, shown in dark blues) in the \textsc{FastPM} mocks since the mass resolution is not high enough to resolve them. The top panel shows the $f_{\rm NL} = 0$ case, and the bottom panel shows the $f_{\rm NL} = 100$ one.} 
 \label{fig:3a_pk_unit_pk_fastpm}
\end{figure}

\autoref{fig:3a_pk_unit_pk_fastpm} shows the power spectrum of \textsc{FastPM} compared to that from the \textsc{PNG-UNIT}. The lines are the power spectrum computed for the \textsc{PNG-UNIT} for the mass bins from \autoref{tab:mass_bins}. The shaded areas are the mean and the standard deviation of the power spectra obtained from the \textsc{FastPM} mocks using the above procedure.

 There are no halos in the fast mocks for the three lowest-mass bins since the mass resolution is not high enough to resolve them. In the case of the two most massive bins, although we can recover the clustering, we cannot simultaneously recover the abundance. This problem is associated with the mass and bias function differences due to the distinct halo finders used (see \autoref{fig:2_halo_mass_function}). In addition, the power spectra for the three most massive bins in the \textsc{PNG-UNIT} are very noisy since the number density is very low $(n<10^{-5} \, h^3 \, {\rm Mpc}^{-3})$. 
\begin{table*}
\centering
\caption{PNG parameters derived for halos in different mass bins.}

\begin{tabular}{@{}cc||c||ccccc@{}}
\hline
Bin & Mass Range            &$b_\phi f_{\rm NL }$            & $b_\phi f_{\rm NL }$  & $f_{\rm NL }\,  (p=1) $ &  $\rho$   & $f_{\rm NL }\,  (p=1) $ & $p$ \\ 
    & $(h^{-1} \, M_\odot )$&$\sigma_{S,\; {\rm UNIT}} $     & $\sigma_{F+M}$      & $\sigma_S$                &           & $\sigma_{F+M}$         & $\sigma_{F+M}$ \\ \hline \hline
  
0   & $(2,5]\times 10^{10}$  &   $16\pm55$      & $-37.9\pm7.0$    & $30\pm 130$      &  $0.703$      & $94\pm10$          & $0.993\pm0.021$ \\ \hline
1   & $(5,10]\times 10^{10}$ &   $4\pm54$       & $7.4\pm9.1$      & $-90\pm 400$     &  $0.703$      & $-55\pm45$         & $0.938\pm0.027$ \\ \hline
2   & $(1,2]\times 10^{11}$  &   $2\pm63$       & $48\pm12$        & $490 \pm 600$    &  $0.703$      & $409\pm 74$        & $0.892\pm0.035$ \\ \hline
3   & $(2,5]\times 10^{11}$  &   $-8\pm69$      & $67\pm11$        & $140\pm 150 $    &  $0.807$      & $138\pm25$         & $0.946\pm0.034$ \\ \hline
4   & $(5,10]\times 10^{11}$ &   $-15\pm83$     & $128\pm18$       & $157\pm100$      &  $0.767$      & $138\pm18$         & $0.895\pm 0.055$ \\ \hline
5   & $(1,2]\times 10^{12}$  &   $10\pm100$     & $168\pm29$       & $112\pm68$       &  $0.736$      & $111\pm15$         & $0.951\pm0.086$ \\ \hline
6   & $(2,5]\times 10^{12}$  &   $-50\pm110$    & $244\pm26$       & $79\pm52$        &  $0.706$      & $104\pm11$         & $0.973\pm0.077$ \\ \hline
7   & $(5,10]\times 10^{12}$ &   $-110\pm140$   & $307\pm43$       & $60\pm46$        &  $0.665$      & $85\pm12$          & $1.16\pm0.13$ \\ \hline
8   & $(1,2]\times 10^{13}$  &   $-230\pm160$   & $476\pm90$       & $54\pm39$        &  $0.635$      & $93\pm13$          & $1.11\pm0.27$ \\ \hline
9   & $(2,5]\times 10^{13}$  &   $50\pm260$     & $630\pm120$      & $89\pm43$        &  $0.605$      & $88\pm16$          & $1.26\pm0.35$ \\ \hline
\textcolor{gray}{10}  & \textcolor{gray}{$(5,10]\times 10^{13}$} &  \textcolor{gray}{$-2138\pm940$}   & \textcolor{gray}{$1020\pm290$}     & \textcolor{gray}{$-58\pm57$}    & \textcolor{gray}{$0.564$}    & \textcolor{gray}{$85\pm26$}          & \textcolor{gray}{$1.53\pm0.93$} \\ \hline
\textcolor{gray}{11}  & \textcolor{gray}{$(1,2]\times 10^{14}$}  &  \textcolor{gray}{$-1796\pm1800$}  & \textcolor{gray}{$880\pm 600$}     & \textcolor{gray}{$-70\pm 100$}  & \textcolor{gray}{$0.534$}    & \textcolor{gray}{$48\pm 36$}         & \textcolor{gray}{$3.8\pm 1.9$} \\ \hline
\textcolor{gray}{12}  & \textcolor{gray}{$(2,5]\times 10^{14}$}  &  \textcolor{gray}{$-548\pm1800$}   & \textcolor{gray}{$-4230\pm 740$}   & \textcolor{gray}{$-15\pm 79$}   &  \textcolor{gray}{$0.504$}    & \textcolor{gray}{$-145\pm 27$}       & \textcolor{gray}{$22.4\pm2.4 $} \\ \hline

\end{tabular}
\tablefoot{The first column shows the ID of each bin. The second column presents the mass range for the halos as taken from the \textsc{PNG-UNIT} (Section\ref{sec:mass_bins}). In the third column, we show the measurements we get for the product $b_\phi f_{\rm NL}$ using the standard errors ($\sigma_S(P(k))$) for the Gaussian \textsc{UNIT} simulation. In the fourth column, we present the results of $b_\phi f_{\rm NL}$ after applying the matching between the \textsc{PNG-UNIT} and the \textsc{UNIT} simulations (Section\ref{sec:bphifnl_vs_b1}). The fifth column provides the \fnl values obtained from the \textsc{PNG-UNIT} simulation using standard errors and assuming the universality relation, $p=1$, in \autoref{eq:bphi_p}. The sixth column shows the Pearson correlation coefficient, $\rho$, assumed for each halo mass bin. These values have been obtained after a linear regression of the quantities measured for bins 3--10, as described in Section \ref{sec:matching}. For bins 0--2, we use instead the mean of the measurements of $\rho$, which is $\bar{\rho}=0.703$, as discussed in the same section. In the seventh column, we present the results when applying the reduced errors derived from the \textsc{FastPM} mocks and matching them with the Gaussian simulation (Section\ref{sec:fnl_vs_m}). The last column displays the values of $p$ obtained from \autoref{eq:bphi_p}, which are used to correct the value of \fnl in the fifth column to the input value (Section\ref{sec:p_vs_m}). We find that the error estimates are not robust for the bins in grey and may lead to biases in the conclusions about $b_\phi$. Therefore, as discussed in Section \ref{sec:bphifnl_vs_b1}, they are not considered for subsequent analyses.}
\
\label{tab:mass_bins}
\end{table*}

\subsection{Variance estimation from fast mocks for fixed simulations}\label{sec:variance_estimation}

We focus now on the computation of the covariance matrix for the power spectrum. In the linear regime, the covariance matrix can be assumed to be nearly diagonal. Therefore, the expected variance for a Gaussian density field can be computed following the expression \citep{Feldman_1994}:
\begin{equation}
    \sigma_S^2(k) = \frac{4\pi^2}{V k^2 \Delta k} \left(P(k) + \frac{1}{\bar{n}}\right)^2\; ,
    \label{eq:pk_Gaussian_variance}
\end{equation}
\noindent
where $P(k)$ is the power spectrum, $\Delta k$  is the width of the power spectrum bins, $\bar{n}$ is the mean number density of tracers, and $V$ is the simulation volume. We refer to this quantity as $\sigma_S(P(k))$, the standard error.

As the initial conditions of the simulations were fixed to their expectation value (\S\ref{sec:simulations}), a considerable suppression is expected with respect to \autoref{eq:pk_Gaussian_variance}, particularly relevant for the largest scales \citep{Angulo_2016,Villaescusa_Navarro_2018,Avila_2022, Maion_2022}.

Our approach is to estimate directly $\sigma(P(k))$ as the standard deviation of the \textsc{FastPM} mocks and apply them for the parameter estimation using the full N-body simulation. However, this quantity is still noisy due to the limited number of mocks available to calculate it. Moreover, we still have a problem obtaining this quantity for bins beyond the resolution limit of the \textsc{FastPM} mocks.

Therefore, we propose fitting the ratio $\sigma(P(k))/P(k)$ with a smooth function in order to get a less noisy estimate of the errors. For that, we measure  $\sigma(P(k))/P(k)$ directly from the \textsc{FastPM} mocks (for the mass bins 3-10 from \autoref{tab:mass_bins} whose power spectra are consistent with the ones coming from the \textsc{PNG-UNIT}) and fit it to the following ansatz that we find convenient after several trials of different functional forms: 
\begin{equation}
    \frac{\sigma((P(k))}{P(k)} = f(k; a,b,c) = a \left(1+\frac{1}{1+b \exp{(-c k})} \right) \; .
    \label{eq:sigpk_over_pk}
\end{equation}

\begin{figure}
 \includegraphics[width=\columnwidth]{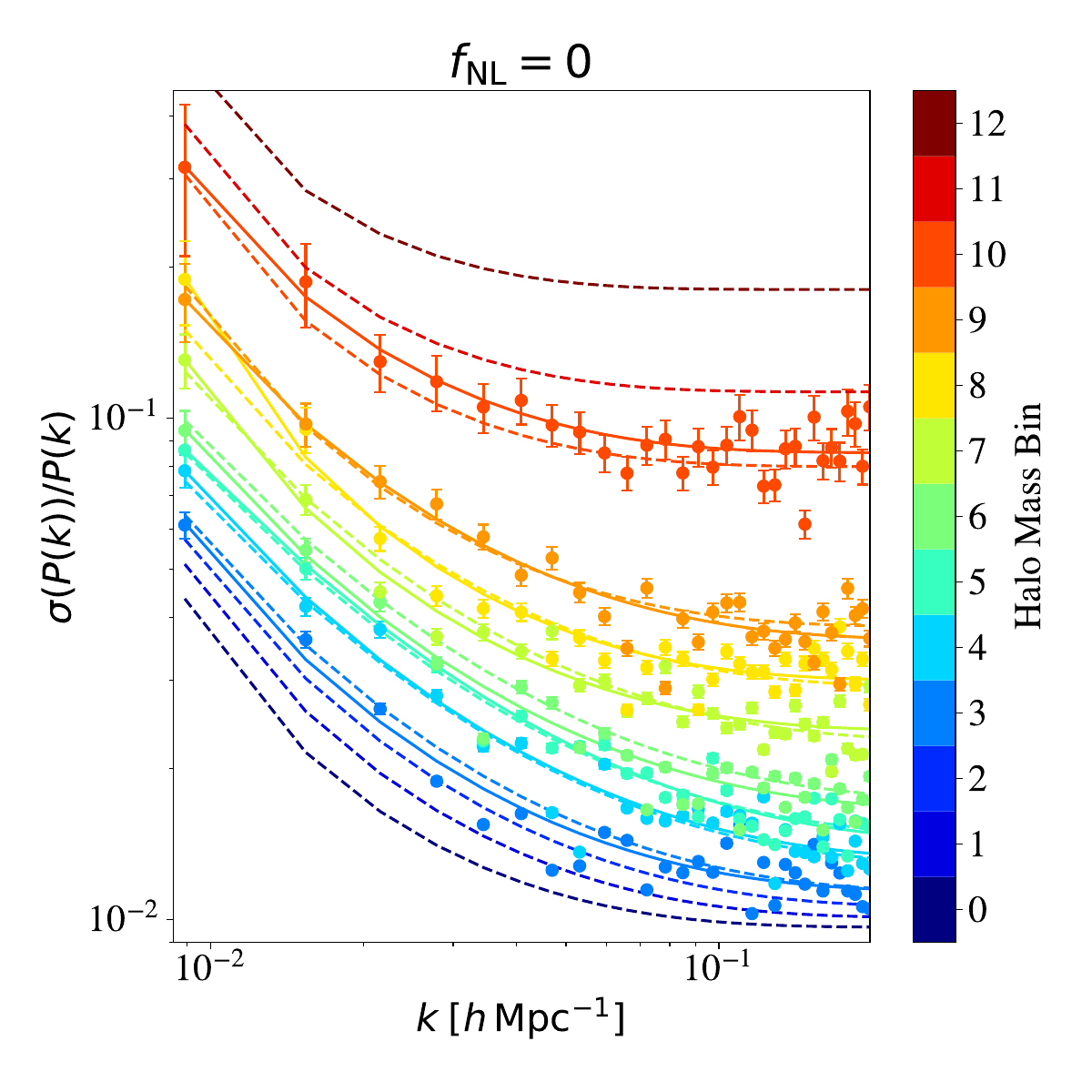}
 \includegraphics[width=\columnwidth]{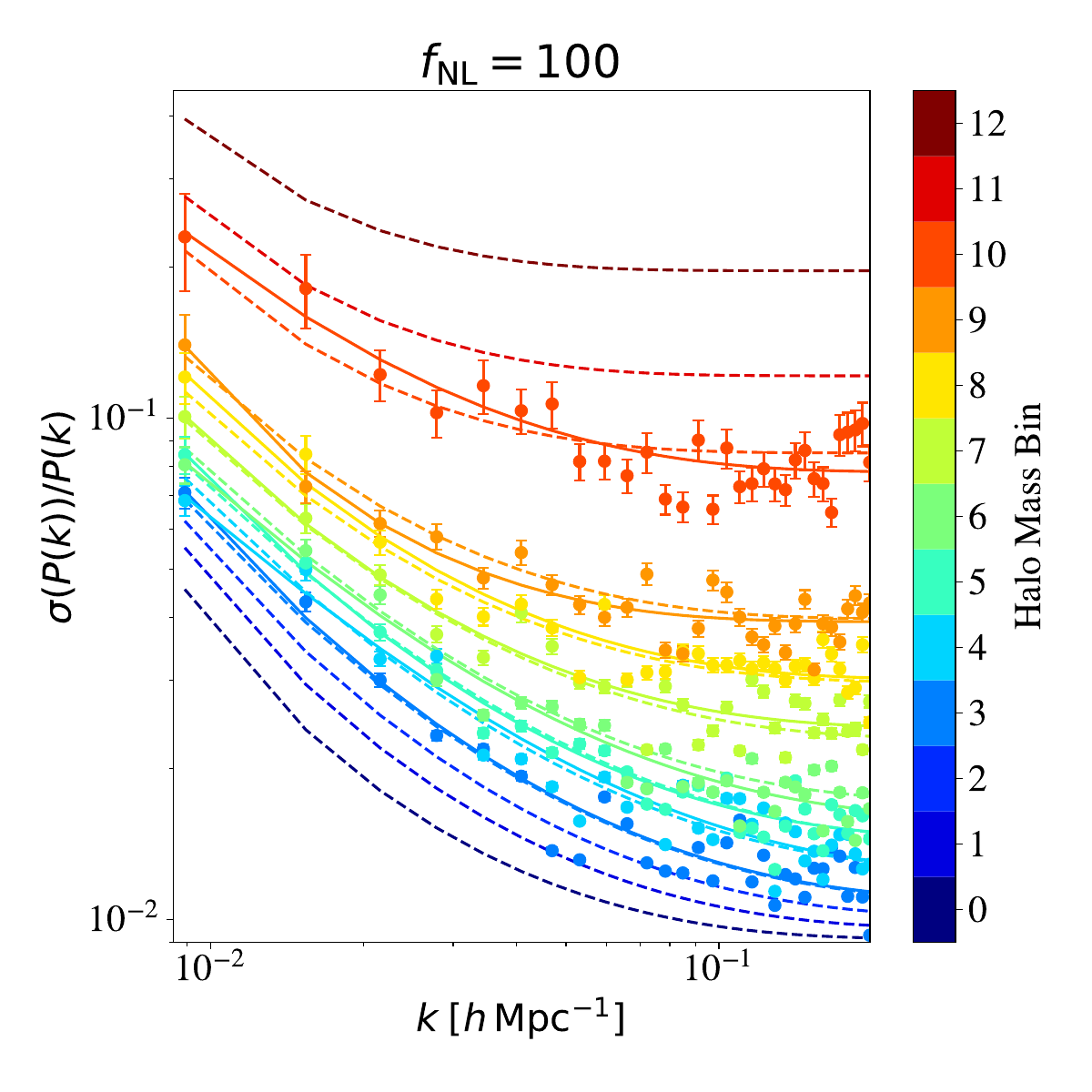}
 \caption{Variance over mean power spectrum for each halo mass bin (indicated by colour with the mass ranges defined in \autoref{tab:mass_bins}). Markers with error bars show the data taken directly from the \textsc{FastPM} mocks, while solid lines show results obtained by fitting these curves to \autoref{eq:sigpk_over_pk}. The extrapolation to each mass bin as described in Section \ref{sec:variance_estimation} is shown in dashed lines.}
 \label{fig:4a_variance_from_fastpm}
\end{figure}

In \autoref{fig:4a_variance_from_fastpm}, we  show the fits of the \textsc{FastPM} mocks to the previous expression  in solid lines. To extrapolate $\sigma(P(k))/P(k)$ beyond the mass resolution of the fast mocks, we fitted the three parameters $a$, $b$ and $c$  from  \autoref{eq:sigpk_over_pk}  we obtained as a function of the mass. For parameter $a$, we use an exponential function; for parameters $b$ and $c$, we use a polynomial of order two as a function of $\log_{10} M_{\rm halo}$. The extrapolations are shown in \autoref{fig:4a_variance_from_fastpm} as dashed lines.

We will now adopt by default the errors coming from the fit to \autoref{eq:sigpk_over_pk} for bins 3-10 and the extrapolation for bins 0-2 and 11-12. Nevertheless, we checked the impact of this assumption, as detailed in Section \ref{sec:bphi_vs_kmax}. 

\subsection{Reducing the errors using fixed-and-matched simulations}\label{sec:matching}

When two simulations share the same set of random phases in their initial conditions, their cosmic variance is highly correlated even if they have different cosmologies. Here, we discuss how we take advantage of the correlated noise from two simulations with different \fnl to constrain the PNG bias parameters with high precision. This noise reduction technique is what we call `matching the simulations'.

The initial conditions for an N-body simulation are a realisation of a Gaussian random field generated from the power spectrum of primordial perturbations. There are two degrees of freedom: the amplitude of the perturbations, whose expected value is determined by the power spectrum, and the phase of those perturbations, which are randomly uniformly distributed between $0$ and $2\pi$. This stochasticity in the generation of the initial conditions (ICs), usually called the cosmic variance, leads to some degree of noise when measuring quantities from simulations in the same way as we expect it to be in the real Universe. The level of noise can be suppressed, for example, by fixing the amplitudes of the modes to their expectation value \citep{Angulo_2016}. 

The idea presented in \citet{Avila_2022} is that if the initial conditions of two simulations share the same stochastic part, that is, the same random phases and the same deviations in the amplitude from their corresponding average (in case they do not have fixed ICs), then the cosmic variance is expected to be highly correlated. Hence, we could cancel out much of the noise affecting measurements by matching that simulation with an existing one with the same ICs but a different cosmology.

We applied this in \citet{Avila_2022} for measuring \fnl, where only three ingredients were needed for the measurement: the value of \fnl obtained from the Gaussian simulation, the same quantity from the non-Gaussian simulation and the correlation coefficient $\rho$ between both simulations. In that study, we used a set of full N-body simulations with matched ICs (i.e. the same stochastic part) to measure $\rho$.

In this study, we adopt an alternative strategy by leveraging only two simulations, \textsc{UNIT} and \textsc{PNG-UNIT}, along with a suite of computationally efficient mocks to estimate $\rho$. This approach circumvents the need for an extensive series of full N-body mocks, which would be impractically demanding on computational resources at this mass resolution. More precisely, we determine $b_\phi f_{\rm NL}$ for each bin within each \textsc{FastPM} realisation  and then calculate the Pearson correlation coefficient to obtain $\rho$.

We find that the measurements of $b_\phi f_{\rm NL}$ correlate more when the mass of the halos is lower. This finding may be because the lower the mean mass of the bins, the higher the number of halos included, and therefore, the less noisy the power spectrum measurement.  We performed a linear fit of this $\rho$ as a function of the mass to reduce the noise associated with the measurement of this quantity\footnote{The values of $\rho$ measured for bins 3--9 are: $0.852$, $0.763$, $0.614$, $0.781$, $0.748$, $0.468$, $0.696$}. The results of the linear fit are reported in \autoref{tab:mass_bins}. However, studying the evolution of the correlation with mass in depth is beyond the scope of this paper.  

Thus, we are left with the question of obtaining the correlation coefficients for those bins where we do not have \textsc{FastPM} mocks. One possibility is to use the values obtained from the linear regression as a function of bin mass. This regression would give us a correlation coefficient of $\rho \sim 0.9$ for the less massive halo bins (i.e. bins 0-2).  While this does not affect the measurement of \bp, it is critical for the error bars. However, we have no guarantee that this relation of $\rho$ with the mass would saturate at some point. Therefore, we have chosen a more conservative approach (as it results in larger reported uncertainties) and assume for the bins 0--2 the average $\rho$ that we have calculated from the \textsc{FastPM} mocks (i.e. bins 3-10), giving for those bins $\rho=0.703$.

In addition,  we propose to apply this to constrain directly the product $b_\phi f_{\rm NL}$ (instead of just \fnl as done in \citet{Avila_2022}), which is the quantity we are sensitive to. To this end, we construct this new estimator for $b_\phi f_{\rm NL}$, which takes into account the matching between both simulations:

\begin{equation}
\Delta b_\phi f_{\rm NL} = [b_\phi f_{\rm NL}]_{100} - [b_\phi f_{\rm NL}]_{0}\; ,
    \label{eq:delta_bphifnl}
\end{equation}
where $[b_\phi f_{\rm NL}]_{100}$ is the quantity measured in the non-Gaussian simulation with $f_{\rm NL} = 100$ and $[b_\phi f_{\rm NL}]_{0}$ corresponds to the Gaussian one.  Given that the noise in $[b_\phi f_{\rm NL}]_{100}$ and $[b_\phi f_{\rm NL}]_{0}$ is highly correlated, it is expected that the average $\Delta b_\phi f_{\rm NL}$ would converge to the mean much faster than the average of $[b_\phi f_{\rm NL}]_{100}$ or $[b_\phi f_{\rm NL}]_{0}$ alone, since we are cancelling some part of the noise. The variance of $\Delta b_\phi f_{\rm NL}$ is given by
\begin{equation}
\begin{aligned}
\sigma^2(\Delta b_\phi f_{\rm NL}) &= \sigma^2([b_\phi f_{\rm NL}]_{100}) +  \sigma^2([b_\phi f_{\rm NL}]_{0}) \\
 &- 2 \rho \sigma([b_\phi f_{\rm NL}]_{100}) \sigma([b_\phi f_{\rm NL}]_{0})\; ,
\end{aligned}
    \label{eq:sig_delta_bphifnl}
\end{equation}
where $\sigma([b_\phi f_{\rm NL}]_{100})$ is the error of the quantity measured in the non-Gaussian simulation and $\sigma([b_\phi f_{\rm NL}]_{0})$ corresponds to the Gaussian one, and $\rho$ is the correlation coefficient between the two measured using the \textsc{FastPM} mocks. 

While, in principle, this method can be applied to any two values of \fnl, the advantage of choosing one value to be $f_{\rm NL} = 0$ is that $\langle \Delta b_\phi f_{\rm NL} \rangle = \langle   [b_\phi f_{\rm NL}]_{100} \rangle $. From now on, when we apply this matching, we label the errors as $\sigma_{S+M}$ if we use \autoref{eq:pk_Gaussian_variance} for the variance of the power spectrum and $\sigma_{F+M}$ if we use the method described in Section \ref{sec:variance_estimation}.

\subsection{Variance estimation methods}\label{sec:sigma_defs}
In the previous sections, we discuss how to reduce the noise in the measurements of $b_\phi f_{\rm NL}$ by taking advantage of the fixed-and-matched initial conditions of the \textsc{PNG-UNIT}.   We compare our results with those derived when no noise reduction techniques are employed to ensure that we are not introducing any bias when using these techniques.

We defined the following errors for the power spectrum, $\sigma(P(k))$:
\begin{itemize}
    \item Standard errors, $\sigma_S$, come from \autoref{eq:pk_Gaussian_variance} with only one simulation used to measure $b_\phi f_{\rm NL}$. This error is not correlated with the other \fnl simulation. This method is expected to provide the largest estimate of the error.
    \item Fixed errors, $\sigma_F$, are obtained from the best-fit $\sigma(P(k))$ (\autoref{eq:sigpk_over_pk}) measured from the \textsc{FastPM} mocks for bins 3--10 in halo mass. For bins 0-2 and 10-12, we do not recover the clustering nor the abundances with \textsc{FastPM} mocks simultaneously, so we need to perform an extrapolation for those mass bins (Section \ref{sec:matching}). To obtain these errors, we do not correlate $f_{\rm NL}=0$ with $f_{\rm NL}=100$.
\end{itemize}
In addition, we have errors that take into account the noise reduction due to the fixed-and-matched initial conditions (\autoref{eq:sig_delta_bphifnl}):
\begin{itemize}
    \item `standard + matched' error, $\sigma_{S+M}$, where the standard error $\sigma_S$ for the power spectrum from the simulations with $f_{\rm NL}=0$ is correlated with the ones with $f_{\rm NL} = 100$ using the method described in Section \ref{sec:matching};
    \item `fixed + matched' errors, $\sigma_{F+M}$, where we assume the $\sigma_{F}$ errors for the power spectrum and correlate matched simulations with different \fnl. We consider these errors for the baseline analysis.
\end{itemize}

In Section \ref{sec:bphi_vs_fastpm}, we compare how these assumptions affect the constraints on \bp and $p$.

\section{Constraining the PNG bias parameters for mass-selected halos}\label{sec:results}

Given the input value of \fnl used to generate the initial conditions for the \textsc{PNG-UNIT}, we can measure directly \bp from the scale-dependent bias (\autoref{eq:power_spectrum_scale_dependent_bias}). In this section, we discuss how we constrain \bp for mass-selected halos using the suppressed variance from  the matching technique (\S\ref{sec:matching}).

First, we check how well \autoref{eq:power_spectrum_scale_dependent_bias} describes the clustering of the dark matter halos in the \textsc{PNG-UNIT} under the standard assumption that \bp is given by the universality relation (i.e. $p=1$). We measure the non-Gaussian signal $b_\phi f_{\rm NL}$ in our simulations (\S\ref{sec:bphifnl_vs_b1}), and by assuming the universality relation, we report our results for \fnl (\S\ref{sec:fnl_vs_m}). Finally, using the input values of \fnl,  we directly measure $p$ (\S\ref{sec:p_vs_m}). In \autoref{tab:mass_bins}, we summarise the main results from this section. 
\subsection{Parameter estimation and analysis choices}\label{sec:choices}
Here, we describe the pipeline applied for parameter estimation common to all our analyses. We study the impact of varying these assumptions in Section \ref{sec:robustness_tests}.
 
To obtain the PNG parameters, we performed Markov chain Monte Carlo (MCMC) using the Python package \textsc{emcee}\footnote{\url{https://emcee.readthedocs.io/en/stable/}} \citep{Foreman_Mackey_2013}. 
All the cosmological parameters are fixed except $b_1$ and $b_\phi f_{\rm NL}$ for which we set flat priors over a wide range ($[0,10]$ for $b_1$ and $[-10000,10000]$ for $b_\phi f_{\rm NL}$). We assume a Gaussian likelihood with $\chi^2$  given by
\begin{equation}
    \chi^2 = \sum^{N_{k}}_{i=0} \frac{\left(P_{\rm halos}(k_i) - P_{\rm model}(k_i; b_1 ;b_\phi f_{\rm NL})\right)^2}{\sigma^2(P(k_i))}\; .
    \label{eq:chi2}
\end{equation}
Here, $P_{\rm model}(k)$ is obtained using \autoref{eq:power_spectrum_scale_dependent_bias} by inputting the matter power spectrum $P_{m,m}(k)$ computed directly from the simulation. In the baseline analysis, we use the fixed+matched errors $\sigma_{F+M}$ defined in Section \ref{sec:sigma_defs} for $\sigma(P(k))$. Moreover, in the subsequent figures, we use solid circles when we are using the best fit to \autoref{eq:sigpk_over_pk} and the mocks, and we use empty circles when we performed extrapolation (i.e. for bins 0-2). We then obtain $\sigma(b_\phi f_{\rm NL})$ from the posterior distribution after marginalising over $b_1$. The model described by \autoref{eq:power_spectrum_scale_dependent_bias} is only accurate in the purely linear regime.

Due to the limitation of our model to the linear regime, we need to drop the modes associated with higher-order nonlinear terms in the power spectrum and consider only the linear modes by applying a scale cut. We choose to set $k_{\rm max} = 0.1 \; h\, {\rm Mpc}^{-1}$, as the next-to-leading order terms become important beyond this scale. In Section \ref{sec:robustness_tests}, we discuss this issue and study its impact on the \bp and \fnl constraints, where we find that variations of $50\%$ in $k_{\rm max}$ compared to our fiducial choice have a negligible effect.


\subsection{Constraints on $b_{\phi} f_{\rm NL}$}\label{sec:bphifnl_vs_b1}

Following the pipeline described above, we fix all the cosmological parameters for our analysis, except $b_1$ and $b_\phi f_{\rm NL}$ in \autoref{eq:power_spectrum_scale_dependent_bias}, and then search for the best-fit parameters.

\begin{figure}
 \includegraphics[width=\columnwidth]{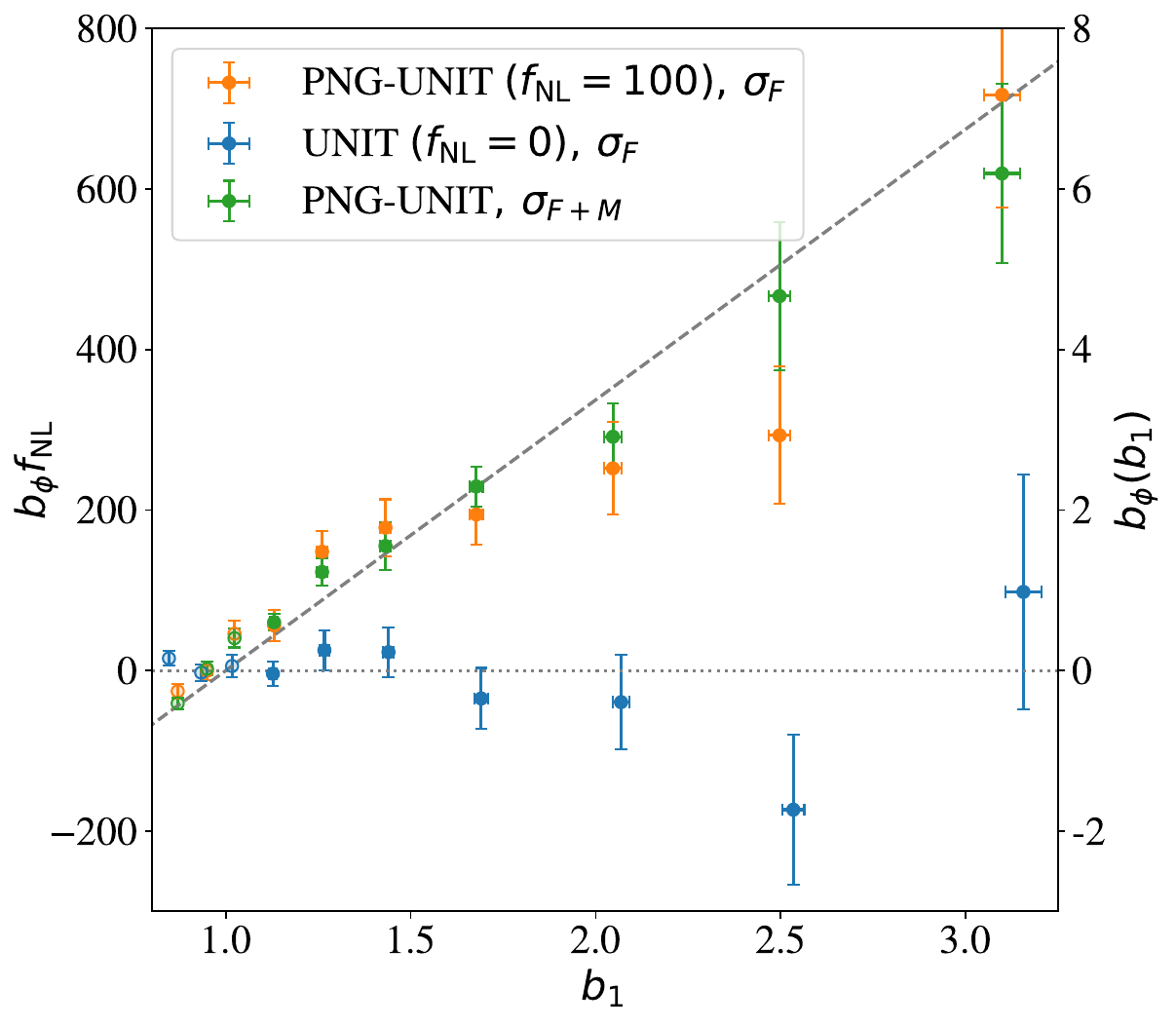}
 \caption{Constraints on $b_\phi f_{\rm NL}$ as a function of $b_1$. Blue and orange symbols show results for the Gaussian and the non-Gaussian simulations, respectively, assuming that the $P(k)$ variance is computed from the \textsc{FastPM} mocks (see \S\ref{sec:variance_estimation}). By subtracting both, we obtain the green points. Thanks to the so-called matching technique, these points have smaller uncertainties (see \autoref{eq:delta_bphifnl} and \autoref{eq:sig_delta_bphifnl} in \S\ref{sec:matching}).  Grey lines indicate the expected value by assuming the universality relation, with the dotted line for $f_{\rm NL} = 0$ and the dashed line for $f_{\rm NL} = 100$. In solid circles, we use the best fit to \autoref{eq:sigpk_over_pk} for estimating the power spectrum variance; in rings, we use the extrapolation method. Here, mass bins 0-9 are shown.}
 \label{fig:5_bphifnl_vs_b1}
\end{figure}

We show the results in the third column of \autoref{tab:mass_bins} and in \autoref{fig:5_bphifnl_vs_b1},  where the blue and orange symbols represent the best-fit values for the Gaussian and non-Gaussian simulations, respectively, and the dashed and dotted lines indicate the expected values for each case. The error bars show the $68\%$ confidence interval obtained from our fits for $b_{\phi} f_{\rm NL}$. For the blue and orange data points, we have used the variances obtained from the fits to the \textsc{FastPM} mocks (see \S\ref{sec:variance_estimation}), while for the green symbols, we have used the fixed-and-matched technique (see \S\ref{sec:matching}). We expected to obtain exactly zero for the Gaussian simulation, while for the non-Gaussian one, the reference we compare to is the universality relation prediction (i.e. $p=1$). The grey lines show these expectations.  For halos with $b_1 > 1$, we clearly measure $b_\phi f_{\rm NL}$. When $b_1\sim1$, for the most popular assumption of $p=1$ in \autoref{eq:bphi_p}, $b_\phi$  is nearly zero and the non-Gaussian signal is masked. In the same figure, we also observe fluctuations for the $f_{\rm NL}=0$ simulation around zero, which is expected due to the intrinsic cosmic variance of the simulation. 

Since both simulations originate from the same random phases, we expect the measurements of $b_\phi f_{\rm NL}$ to be highly correlated (\S\ref{sec:matching}). Therefore, if we find perturbation in the Gaussian simulation ($f_{\rm NL}=0$) that lowers the measured $b_\phi f_{\rm NL}$, an equivalent perturbation is expected to appear in the non-Gaussian ($f_{\rm NL}=100$) simulation, as we see in \autoref{fig:5_bphifnl_vs_b1} indeed.

We apply Equations \ref{eq:delta_bphifnl} and \ref{eq:sig_delta_bphifnl} to compute the matched data points from both simulations, obtaining the green points in \autoref{fig:5_bphifnl_vs_b1}.  We expect that these results have less noise than the individual cases of the simulations because of the correlations in the initial conditions. We find the green points to be closer to the prediction from the universality relation ($p=1$). While this is true for $M_{\rm halo}> 1\times 10^{12} h^{-1}M_\odot$, we find that the measured $p$ is below that of the universality relation for masses with $M_{\rm halo}< 1\times 10^{12} h^{-1}M_\odot$, with a significance that reaches $3\sigma$ for bin 2.  

Regarding the mass range from  $M_{\rm halo}= 5\times 10^{13} h^{-1}M_\odot$ up to $5\times 10^{14} h^{-1}M_\odot$ (bins 10-12 in \autoref{tab:mass_bins}, not shown from \autoref{fig:5_bphifnl_vs_b1} on), their power spectra exhibit a very low signal-to-noise ratio $({\rm S/N}<5)$ due to their low number densities $(n_{\rm halos} \ll 10^{-4} \, h^3 \, {\rm Mpc}^{-3})$.  As we discussed in Section \ref{sec:variance_estimation}, to estimate the covariance for the power spectra, we considered two cases: applying \autoref{eq:pk_Gaussian_variance} or the extrapolation of \autoref{eq:sigpk_over_pk}. When considering $\sigma_S(P(k))$, a more detailed analysis of these bins reveals that considering a super-Poisson contribution to the shotnoise ($\propto (1+\alpha)/n_g$) is necessary to have an accurate estimation of the variance of the power spectrum, $\sigma (P(k))$. At the same time, this extra noise, parameterised by $\alpha$, is consistent with $\alpha =0$ for the rest of the bins (0--9), which is our assumed baseline.  We leave for future analysis the usage of non-Poisson errors as our analysis baseline. Thus, in what follows, we only analyse halos with masses $2\times 10^{10} < M_{\rm halo}(h^{-1}M_\odot)\leq 5\times 10^{13}$, which correspond to bins 0 to 9 in \autoref{tab:mass_bins}.


\subsection{Measuring $f_{\rm NL}$ from $b_\phi f_{\rm NL}$ assuming the universality relation} \label{sec:fnl_vs_m}

We cannot constrain both \bp and \fnl simultaneously since they are perfectly degenerate. We can do so, however, by assuming the universality relation with $p=1$. In this way, we extract \fnl from our simulation results. If this relation does not accurately describe the \bp parameter, we expect to recover biased values of \fnl. Hence, we have tested the validity of the universality relation.

\begin{figure}
 \includegraphics[width=\columnwidth]{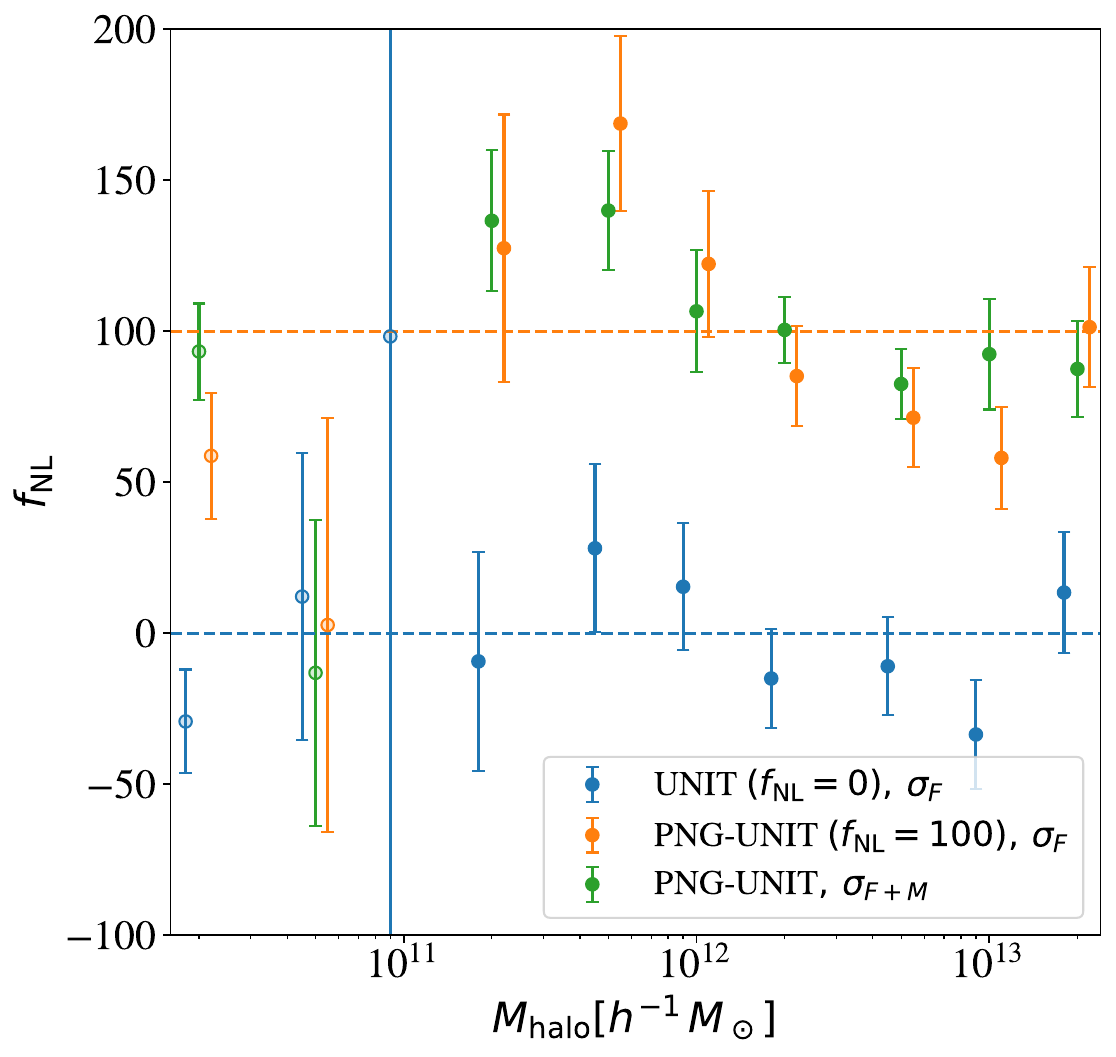}
 \caption{Value of \fnl for each mass bin, assuming \bp is described by \autoref{eq:bphi_p} with $p=1$. Blue and orange symbols show results for the Gaussian and non-Gaussian simulations, respectively. In contrast, green symbols show the matching (difference) between them, where we used \autoref{eq:delta_bphifnl} for obtaining $b_\phi f_{\rm NL}$ and \autoref{eq:sig_delta_bphifnl} for its uncertainty. Dashed lines show the input \fnl values for each simulation. For the mass bins 2 and 3, we have that $b_1 -1 \leq 0.1$ and hence $b_\phi \sim 0$. This value of \bp explains the large error bars compared to the other mass bins.  In solid circles, we use the best fit to \autoref{eq:sigpk_over_pk} for estimating the power spectrum variance; in rings, we use the extrapolation method.}
 \label{fig:6_fnl_vs_m}
\end{figure}

In \autoref{fig:6_fnl_vs_m}, we compare the values of \fnl obtained in this way with the ones set in the initial conditions.  We find that the perturbations that favour a positive value of \fnl in the Gaussian simulation are also present in the non-Gaussian one. The results are shown in blue symbols for the $f_{\rm NL} = 0$ simulation and orange symbols for the $f_{\rm NL} = 100$  simulation. We reduce the noise in \fnl by applying the matching technique. This involves the estimation of $b_\phi f_{\rm NL}$ using \autoref{eq:delta_bphifnl} and its variance through \autoref{eq:sig_delta_bphifnl}. In the Gaussian simulation, the expected value for $b_\phi f_{\rm NL}$ is 0. This means that the expected value of $\Delta b_\phi f_{\rm NL}$ is the same as the  $b_\phi f_{\rm NL}$ of the simulation with $f_{\rm NL}=100$. Then, we divide by $b_\phi$ computed assuming $p=1$. Using this correlation, we determine the combined measurement of \fnl, shown in green symbols in \autoref{fig:6_fnl_vs_m}. 

For masses between $5\times 10^{10}$ and  $2\times 10^{11} \; h^{-1} M_\odot$, the error bars are enlarged because $b_1 \sim 1$ for those bins, so using the universality relation results in $b_\phi \sim 0$. Moreover, a slight deviation from the expected value in $b_\phi f_{\rm NL}$ assuming universality in these mass bins leads to an amplification of this deviation from the input value in \fnl, precisely because $b_ \phi$ is close to $0$. This is what we observe for bins 2 and 3. Nevertheless, considering also the growth of the errorbars, for halos with masses between $M_{\rm halo} =  1\times 10^{11} \; h^{-1} M_\odot$ and $M_{\rm halo} = 1\times 10^{12} \; h^{-1} M_\odot$ we find deviations from the input value, with a significance up to $4\sigma$ in the case of bin 2. These differences, which we have already seen in $b_\phi f_{\rm NL}$ in \autoref{fig:5_bphifnl_vs_b1}, can be interpreted as meaning that the population of halos in that mass bin have \bp that does not follow the universality relation, as we assumed $p=1$ for deriving these measurements of \fnl. Finally, we recover the input value of \fnl  for both simulations for halos with $M_{\rm halo} > 1\times 10^{12} \; h^{-1} M_\odot$ (see \autoref{tab:mass_bins}).


\subsection{p as a function of mass} \label{sec:p_vs_m}

Since we know the input value of \fnl, we can use the deviations of our measurements from that value to find how far the assumption $p=1$ is from the $p$ measured for the simulation halos.

We fix the value of \fnl and derive \bp from the previous constraints on $b_\phi f_{\rm NL}$ we obtained in Section \ref{sec:bphifnl_vs_b1}. We then compute $p$ using \autoref{eq:bphi_p}. We have also checked that we get the same results by fitting $p$ directly and fixing \fnl to its input value in a similar way as done in Section \ref{sec:fnl_vs_m}.

\begin{figure}
 \includegraphics[width=\columnwidth]{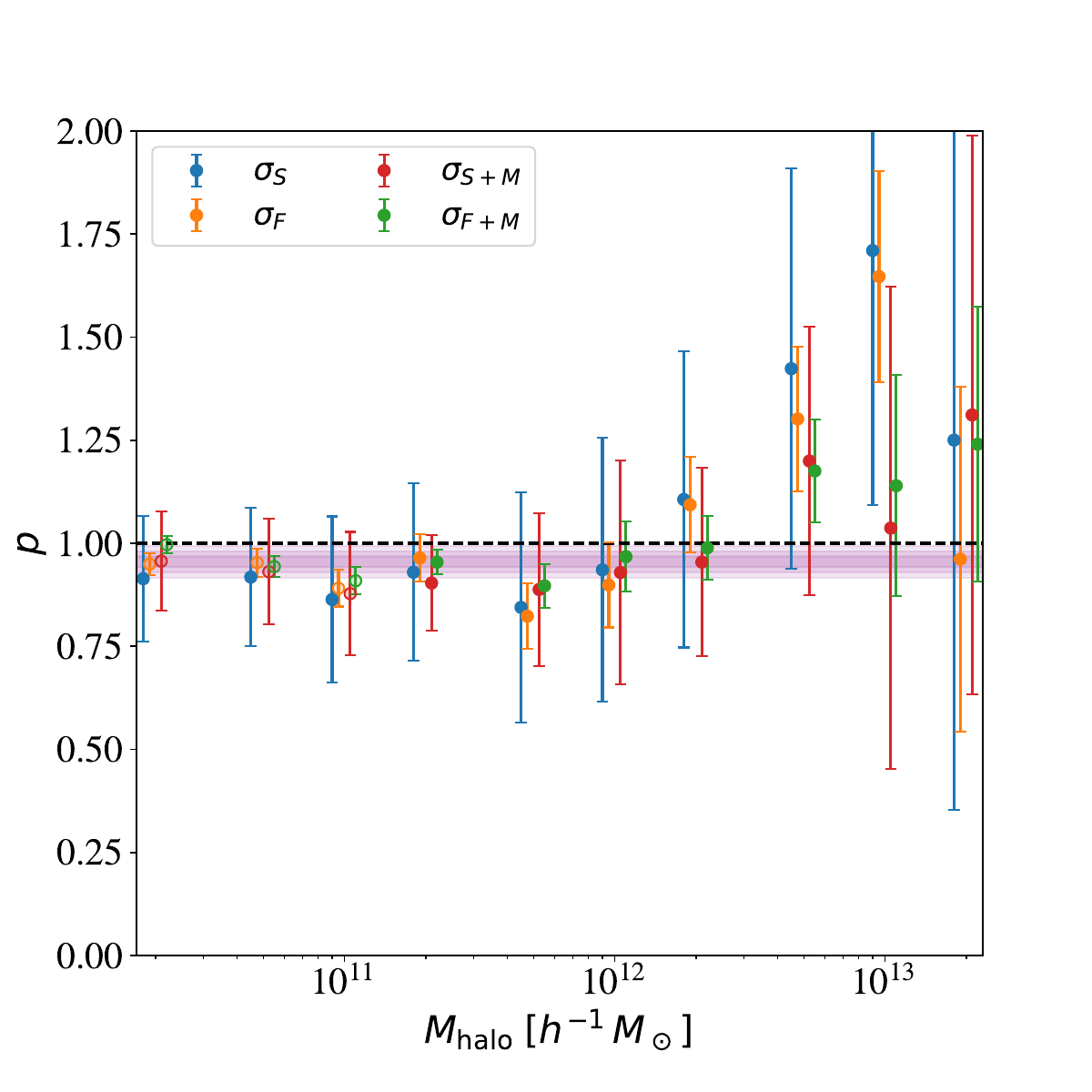}
 \caption{PNG-response parameter $p$ as a function of the halo mass, using different methods to treat variances. Blue and orange symbols show results for the non-Gaussian simulation, using standard variances (\autoref{eq:pk_Gaussian_variance}) and fixed variances, respectively. Green and red symbols show the same after applying the matching with the Gaussian simulation. In solid circles, the fixed variances is estimated by the best fit to the \textsc{FastPM} variances with \autoref{eq:sigpk_over_pk}; in rings, we use the extrapolation method. The shaded areas represent the $1\sigma$, $2\sigma$ and the $3\sigma$ regions (from darker to lighter respectively) around the value of $p$ we find in \autoref{eq:p_result}.}
 \label{fig:7_p_vs_m}
\end{figure}
The results are shown in \autoref{fig:7_p_vs_m}, where we compare the different methods of estimating the variance of the power spectrum and reducing the noise introduced in Section \ref{sec:sigma_defs}. Our methodology is consistent as we recover the same results regardless of the method used for estimating the variance of the power spectrum with or without the matching technique.  The clustering of the  halos with $1\times10^{12}<M_{\rm halo}<2\times10^{13} \; h^{-1} M_\odot$ are aptly described through the universality relation $(p=1)$, given the size of our uncertainties. Concerning the  mass bins with $M_{\rm halo}<5\times10^{11}$, we find a slight deviation from $p=1$, with a significance that reach $>3\sigma$ for halos with masses between $1\times10^{11}$ and $2\times10^{11} \, h^{-1} M_\odot$.  These lower values of $p$ are related to the overprediction of the product $b_\phi f_{\rm NL}$ (\S\ref{sec:bphifnl_vs_b1}). 

We explore the possibility of $p$ being a constant independent of the halo mass. We do this by performing a linear fit. We obtain a slope consistent with zero at $1\sigma$, suggesting that a constant $p$ can indeed be assumed for the halos. By an average weighted by the inverse of the variance, we get:
\begin{equation}\label{eq:p_result}
    p=0.955\pm0.013
\end{equation}
with $\chi^2/{\rm dof} = 1.31$ when we take into account all the bins (0--9), while the  $\chi^2/{\rm dof}$ for $p=1$ is $2.64$. This best-fit value of $p$ with the $1$,$2$ and $3\sigma$ regions  are shown as the shaded region in \autoref{fig:7_p_vs_m}. If we neglect those bins where we used extrapolation (i.e. by removing the mass bins 0-2 from the fits and considering just 3-9), we get:
\begin{equation}
    p=0.950\pm 0.025
\end{equation} 
with a $\chi^2/{\rm dof}$ of $1.33$. 


\section{Convergence tests}\label{sec:sec_convergence}

This section shows how far we can push our results to low masses.  To do so, we test the convergence of the \textsc{PNG-UNIT} against the \textsc{LR-UNIT} simulations with a lower resolution (\S\ref{sec:convergence_lr}). We also demonstrate the robustness of our results by comparing them with an alternative method to measure \bp, using the separate universe approach (\S\ref{sec:convergence_sep}). Finally, we investigate the impact of considering different mass definitions (\S\ref{sec:bphi_vs_mdef}).

\begin{figure}
 \includegraphics[width=\columnwidth]{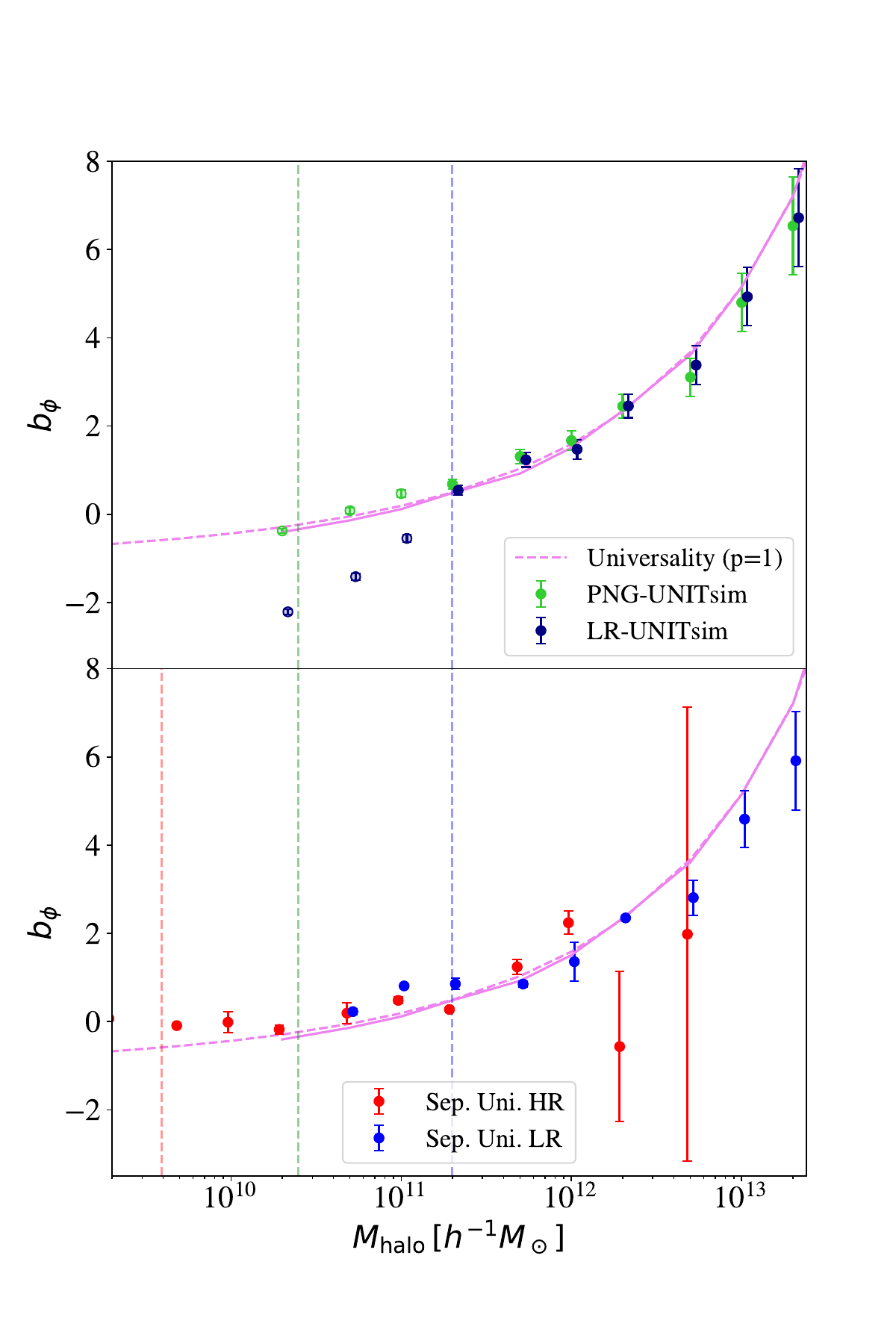}
 \caption{Non-Gaussian bias parameter \bp as a function of the halo mass for the different sets of simulations. The colour of the symbols and the dashed lines represent the mass resolution. Red:  $m_p = 1.5 \times 10^{8} \, [h^{-1}\,M_\odot]$. Green: $m_p = 1.2 \times 10^{9} \, [h^{-1}\,M_\odot]$. Blue-Dark blue: $m_p = 1.0 \times 10^{10} \, [h^{-1}\,M_\odot]$. The dashed vertical lines represent the threshold of  $M_{\rm halo} = 20 m_{\rm part}$ for each simulation. The pink solid line is the universality relation $p=1$ for the $b_1$ values measured in the \textsc{PNG-UNIT}. The dashed pink line extrapolates the universality relation toward lower masses, using $b_1$ coming from \citet{Tinker_2010}.  In solid circles, we use the best fit to \autoref{eq:sigpk_over_pk} for estimating the power spectrum variance; in rings, we use the extrapolation method. Upper panel: Measurements of \bp in non-Gaussian simulations by fitting \autoref{eq:power_spectrum_scale_dependent_bias} (see \S\ref{sec:convergence_lr}). Lower panel: Measurements of \bp using the separate universe technique (see \S\ref{sec:convergence_sep}).}
 \label{fig:bphi_convergence}
\end{figure}

\subsection{Convergence with LR-UNIT simulations}\label{sec:convergence_lr}
First, we investigate the mass convergence of our results by comparing the \textsc{PNG-UNIT} results with those of the lower-resolution \textsc{LR-UNIT} simulations. Our methodology is the same as in Section \ref{sec:bphifnl_vs_b1}, and we assume $\sigma_{F+M}$ errors.

The upper panel of \autoref{fig:bphi_convergence} compares the values of \bp derived from the $4096^3$-particle simulations with those from the \textsc{LR-UNIT} simulations. The \bp values agree for bins with $M_{\rm halo} \gtrsim 20 m_{\rm part, LR}$. Below this threshold, the mass function from the \textsc{LR-UNIT} simulations drops as we cannot resolve such light halos (\autoref{fig:2_halo_mass_function}). This drop in the abundance of halos also affects the clustering, as demonstrated by the sudden change in \bp. Thus, we prove that the limit of $20$ particles is an adequate minimum threshold for the halo mass considered for our simulations.

The universality relation ($p=1$) is shown as a solid pink line in \autoref{fig:bphi_convergence}, assuming the $b_1$ values from the fits of the \textsc{PNG-UNIT} to the corresponding mass resolution.  To extend the relation beyond the mass resolution of the N-body simulations, we have also used the theoretical values for $b_1(M)$ derived using \citet{Tinker_2010} for each bin in halo mass, given the simulation cosmology. This extrapolation is shown as dashed pink lines in the upper panel of \autoref{fig:bphi_convergence}. These two lines are consistent with each other. 

The derived \bp values agree in both resolutions for halos with more than $20$ particles. Although not shown here, we have checked that the results from running \textsc{LR-UNIT} with \textsc{2LPTic} \citep{Crocce_2006, Crocce_2012} initial conditions are consistent with those from \textsc{LR-UNIT} initialised with \textsc{FastPM}. In a forthcoming paper, we study the impact of using different codes for the initial conditions in simulations with primordial non-Gaussianities.

Finally, we measure $p$ from the \textsc{LR-UNIT} simulations by fixing $b_1$ to \citet{Tinker_2010} in \autoref{eq:bphi_p}. We obtain similar results for $p$ using $b_1$ directly obtained from the simulations. Combining all the mass bins with $M_{\rm halo}>20 m_{\rm part}$, we obtain $p=0.959\pm0.014$ for the \textsc{PNG-UNIT} and $p=0.972\pm0.017$ for the \textsc{LR-UNIT}. These values are consistent with \autoref{eq:p_result}.

\subsection{Comparison with separate universe simulations}\label{sec:convergence_sep}

As a further test of the robustness of our results, we compare two alternative methods for measuring \bp: (i) using the scale-dependent bias as we have done throughout this paper and (ii) using the so-called separate universe technique. This latter technique exploits the fact that \bp is defined as the response of the abundance of halos to the presence of a large-scale perturbation (\autoref{eq:bphi_partial_As}). 

The separate universe technique consists of computing \bp as the numerical derivative of two simulations with the same initial conditions but with a slightly different $\mathcal{A}_s$, or equivalently, $\sigma_8$. The initial conditions for these simulations are Gaussian; thus, these simulations do not show the effect of the scale-dependent bias. Following a similar procedure as done in  \citet{Barreira_2020, Barreira_2022b}, \bp is obtained as follows:
\begin{equation}
    b_\phi = \frac{b_\phi^{\rm high} + b_\phi^{\rm low}}{2} \, ,
\label{eq:bphi_mean}
\end{equation}
with
\begin{equation}
\begin{aligned}
b_\phi^{\rm high} &= \frac{4}{\delta \mathcal{A}_s  N_h^{\rm fid}} (N_h^{\rm high} - N_h^{\rm fid}),  \\
 b_\phi^{\rm low}&= \frac{4}{\delta \mathcal{A}_s  N_h^{\rm fid}} (N_h^{\rm fid} - N_h^{\rm low}) .
\end{aligned}
    \label{eq:bphi_sep_uni}
\end{equation}
In \autoref{eq:bphi_sep_uni}, $\delta \mathcal{A}_s $ is the variation in the amplitude of the initial power spectrum (which we set to $5\%$), and $N_h$ is the number of halos within a particular mass bin measured in the simulations. For the separate  universe technique, we estimate the error in \bp, $\sigma_{\rm SU}(b_\phi)$, following the same consideration as \citet{Barreira_2020} and \citet{Barreira_2022b}: 
\begin{equation}
   \sigma_{\rm SU}(b_\phi) \sim  | b_\phi^{\rm high} - b_\phi^{\rm low}|.
   \label{eq:err_SU}
\end{equation}

The lower panel of \autoref{fig:bphi_convergence} compares the \bp values obtained from using the separate universe technique (\autoref{eq:bphi_sep_uni}) for two sets of simulations with different mass resolutions. Comparing this method for measuring \bp with respect to the one we have been considering in Section \ref{sec:bphifnl_vs_b1} is a way to check the consistency across the two different methods and whether they hold at low masses.  

The universality relation is displayed as pink lines in \autoref{fig:bphi_convergence}. We used both the linear bias, $b_1$, coming from the measurements on the \textsc{PNG-UNIT} simulation (continuous line) and $b_1$ coming from \citet{Tinker_2010} (dashed line). We do not measure $b_1$ from the separate universe simulations as the box size of the simulations is relatively small ($L_{\rm box} = 67.5\, h^{-1}\,{\rm Mpc}$ and $L_{\rm box} = 250\, h^{-1}\,{\rm Mpc}$ for the high-resolution and low-resolution simulations respectively) and direct measurement of the linear bias through the halo power spectrum may be significantly contaminated by the non-linearities of the small scales ($k_{f,\, {\rm HR}} \sim 0.09 \; h\, {\rm Mpc}^{-1}$ and $k_{f,\, {\rm LR}} \sim 0.02 \; h\, {\rm Mpc}^{-1}$).

The error bars obtained for the separate universe method might not be robust. On the one hand, we find that the high-resolution separate universe simulations do not agree with the low-resolution simulations, as shown in the lower panel of \autoref{fig:bphi_convergence}. The real error bars should, at least, reflect the difference between both. On the other hand, the scatter around the universality relation is much bigger than the error bar reported by this method. This is specially significant for $M_{\rm halo}$ between $1\times 10^{11}\, h^{-1} M_\odot$ and $5\times 10^{11}\, h^{-1} M_\odot$.
 
We estimate a global error in \bp by combining the measurements from both the high and low-resolution simulations, taking the mean value and the absolute difference as the error, similar to what we did to compute $b_\phi$ from the high-As and the low-As simulations in \autoref{eq:bphi_mean} and \autoref{eq:err_SU}. With this procedure, we expect to obtain more reliable errors than those from a single set of simulations (i.e. high or low-resolution). 

Finally, we measure $p$ from the separate universe. We fit \bp to \autoref{eq:bphi_p} using a least-squares method, with $b_1$ coming from \citet{Tinker_2010} for each mass bin. Utilising the error in \bp for the combination of the high- and low-resolution separate universe simulations we have described, we get $p=0.8307\pm0.0022$. This result differs from the universality relation and the $p$ obtained for the \textsc{PNG-UNIT}, \autoref{eq:p_result}. We argue that this suggests a shortcoming in the method used to obtain $\sigma_{\rm SU}(b_\phi)$. The mass bins with the smallest errors drive the separate universe measurement of $p$. Moreover, $\chi^2/{\rm dof} = 519/20$ for the $p$ obtained with the separate universe technique is rather large. 

Ignoring the separate universe error bars, we find $p=1.047\pm0.057$ by not weighting the bins by their $\sigma(b_\phi)$ in the least-squares fitting procedure. If we apply this method to the results coming from \textsc{PNG-UNIT} simulation, we get $p=0.984\pm0.031$. In both cases, the value is consistent with the universality relation, and it is less than $2\sigma$ away from the results shown in \autoref{eq:p_result}.  

Overall, the results in this section suggest that one must be very careful when considering the errors in $b_\phi$ arising from the separate universe technique and consider a larger set of simulations to deal with them statistically. Given the setup considered in this paper, the error bars reported are more robust for \textsc{PNG-UNITsims}, and the underlying true uncertainty seems smaller for \textsc{PNG-UNITsims}.

\subsection{Dependency on the mass definition}\label{sec:bphi_vs_mdef}

If \bp depends on the halo mass only by its dependence on $b_1$, then the relationship between these two quantities should be independent of the mass definition used for halos.

Throughout this work, the halo masses have been obtained by computing the enclosed mass within a spherical overdensity up to $200$ times the critical density $\rho_c$ at a given $z$, $M_{200c}$. Here, we redo the analysis to obtain $p$ using other mass definitions, namely: $M_{500c}$, $M_{200b}$ and $M_{\rm vir}$. The radii used for these mass definitions are determined by the threshold density achieved within the spherical overdensity, which reaches $500$ the critical density  $\rho_c$ for $M_{500c}$ and $200$ times the background density $\rho_b$ for $M_{200b}$. For the virial mass $M_{\rm vir}$, the spherical overdensity is determined by the virial radius $r_{\rm vir}$ \citep{Bryan_1998}. 

We have $M_{200c}\geq M_{500c}$ for a given halo. Thus, if we apply the same mass cuts for defining the bin, we expect that $b_1 (M_{200c}) < b_1(M_{500c})$, as if we compare the $M_{200c}$ of the halos selected in each case, we are taking more massive halos when we apply the cut in $M_{500c}$. However, although we expect that  $b_\phi (M_{200c}) \neq b_\phi(M_{500c})$ if \bp only depends on the mass through $b_1$, we would expect that $p$ does not vary with the mass definition.

\begin{figure} \includegraphics[width=\columnwidth]{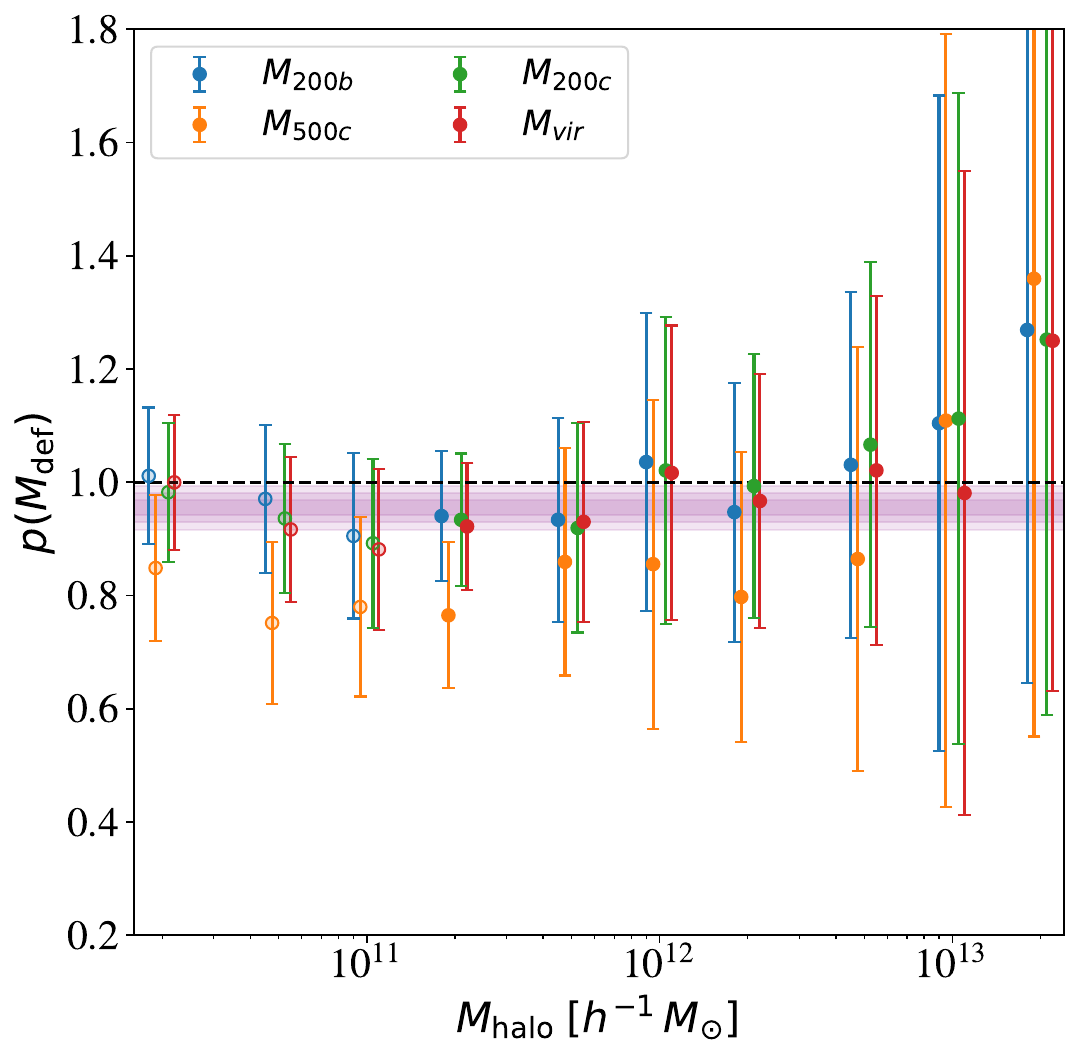}
 \caption{PNG bias parameter $p$ as a function of $M_{\rm halo}$ for different mass definitions. The green symbols represent our choice for the fiducial analysis $(M_{200c})$, while the grey dashed line represents the universality relation ($p=1$). The shaded areas represent the $1\sigma$, $2\sigma$ and the $3\sigma$ regions (from darker to lighter respectively) around the value of $p$ we find in \autoref{eq:p_result}. Masses are shifted for visualisation purposes. } 
 \label{fig:9_mdef}
\end{figure}

We present our findings in \autoref{fig:9_mdef} which compares four mass definitions outputted by \textsc{Rockstar}: our fiducial $M_{200c}$, $M_{500c}$, $M_{200b}$ and $M_{\rm vir}$. There are several things to point out. We find that bins defined according to $M_{200b}$ and $M_{\rm vir}$ are practically equal between them and very similar to $M_{200c}$. Moreover, they have almost the same $b_1$, although this is not shown here. For these three cases, the results are consistent with our findings in Section \ref{sec:p_vs_m}.

However, the most interesting result comes from the bins defined according to $M_{500c}$. We find a systematic suppression of $p$ below the predicted value from the universality relation and compared with the other mass definitions, although the error bars cover $p=1$.  This result suggests that \bp for dark matter halos may not depend solely on the mass through $b_1$ but also on other parameters.  The \bp parameter is defined as how the abundance of the dark-matter halos responds to a large-scale perturbation. However, PNG may not only affect the amplitude of the primordial perturbations (and hence the abundance of the high-mass halos) but also could affect other properties of the halo as suggested by recent works \citep{Sullivan_2023,Lazeyras_2023,Lucie-Smith_2023,Fondi_2023}. Investigating this dependence is beyond the scope of this paper and is left for future research.

We also tested the impact on \bp if we consider the main halos and the subhalos identified by \textsc{Rockstar}. This consideration might be necessary if one runs subhalo abundance matching (SHAM) algorithms to populate the simulation with galaxies. When the substructure is included in the analysis, we find that this induces variations in the linear bias $b_1$ of the mass bins and in $p$. We leave a detailed analysis of this for future work.

\section{Robustness tests}\label{sec:robustness_tests}

Here, we test the impact of the assumptions we have made in our analysis to obtain \bp or $p$. In particular, we study the impact of the scale cut, $k_{\rm max}$ (\S\ref{sec:bphi_vs_kmax}), and the variance estimation methods described in Section \ref{sec:sigma_defs}.
\subsection{Choice of $k_{\rm max}$}\label{sec:bphi_vs_kmax}

\autoref{eq:power_spectrum_scale_dependent_bias} is only valid in the linear regime, thus, we must discard all modes with $k$ greater than a certain $k_{\rm max}$ to avoid introducing non-linear modes our model does not properly describe. As perturbation theory breaks beyond $k_{\rm NL}\sim 0.3  \; h\, {\rm Mpc}^{-1}$ \citep{Desjacques_2018}, $k_{\rm max}$ should be less than this value. Even before reaching this threshold, we are at the risk of incorporating some mildly non-linear modes that our model does not adequately describe.

To determine the optimal $k_{\rm max}$, we must first examine the limits of the linear theory using only the Gaussian simulation.  In this case, \autoref{eq:power_spectrum_scale_dependent_bias} is reduced to $P_{h,h}(k) = b_1^2 P_{m,m}(k)$. The leading correction of the non-linear bias to $P_{h,h}(k)$  has a dependence with $Ak^2$, where $A$ is a nuisance parameter \citep{Lazeyras_2018,Lazeyras_2023}. To establish $k_{\rm max}$, we compare the fits obtained using only $b_1$ with those obtained using $b_1 + A k^2$ after marginalising over the nuisance parameter $A$ as a function of $k_{\rm max}$. We find that the differences in the posterior distributions for $b_1$ obtained using both methods agree for $k_{\rm max} \leq 0.1 \;  h\, {\rm Mpc}^{-1}$ for all mass bins.

\begin{figure} \includegraphics[width=\columnwidth]{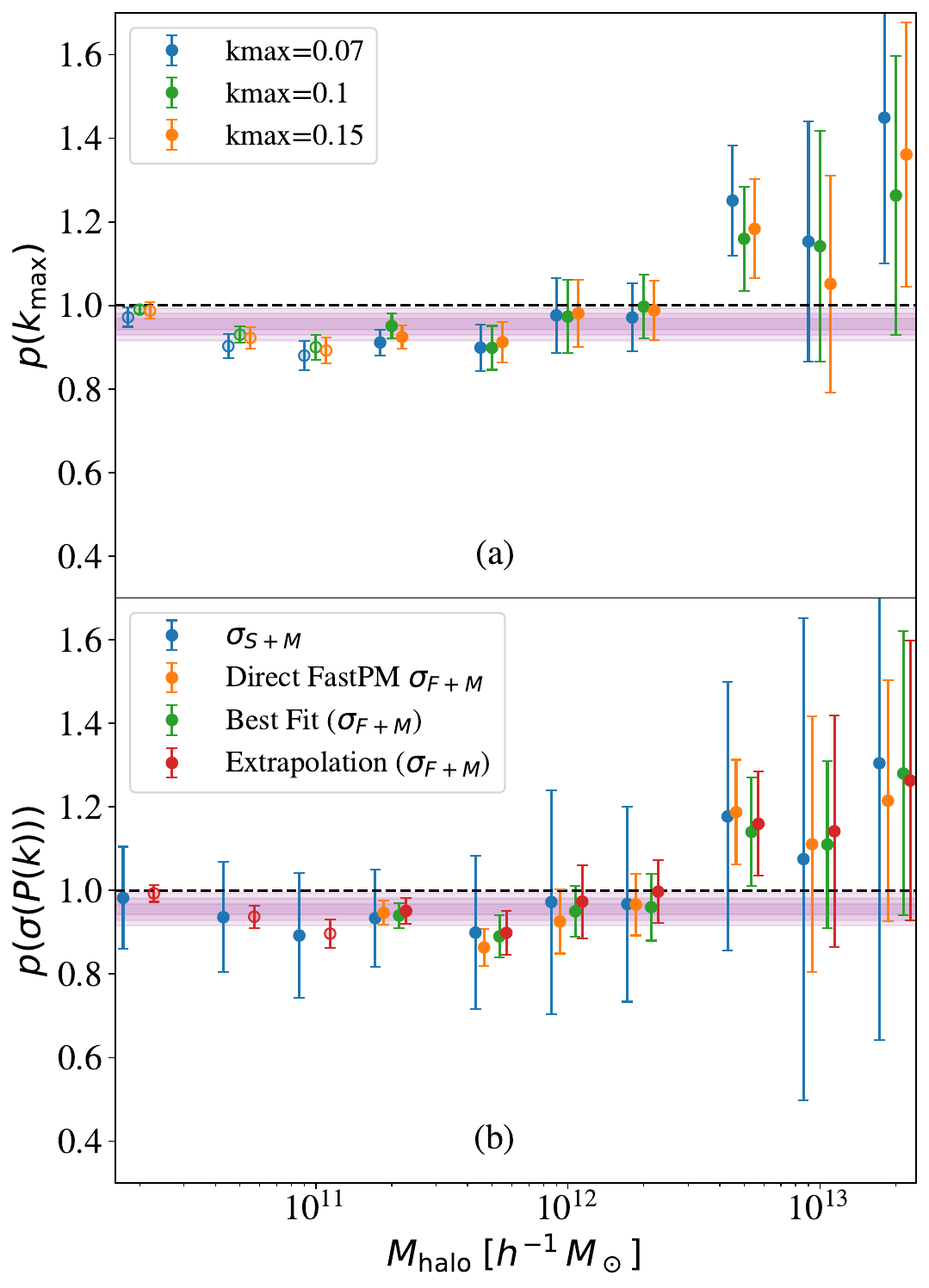}
 \caption{PNG bias parameter $p$ as a function of $M_{\rm halo}$ for different analysis choices. The green symbols represent the fiducial choices, while the grey dashed line represents the universality relation ($p=1$).  In solid circles, we use the best fit to \autoref{eq:sigpk_over_pk} for estimating the power spectrum variance; in rings, we use the extrapolation method. The shaded areas represent the $1\sigma$, $2\sigma$ and the $3\sigma$ regions (from darker to lighter respectively) around the value of $p$ we find in \autoref{eq:p_result}. Masses are shifted for visualisation purposes. Panel \textbf{(a)}: variation in $p$ with changes  $k_{\rm max}$ (see \S\ref{sec:bphi_vs_kmax}). Panel \textbf{(b)}: constraints on $p$ by considering different methods to compute $\sigma(P(k))$, namely: standard $\sigma_S(P(k))$ (blue), direct fit to \textsc{FastPM} mocks (orange), best-fit to \autoref{eq:sigpk_over_pk} (green) and extrapolation of \autoref{eq:sigpk_over_pk} to the mass bins (red) (see \S\ref{sec:bphi_vs_fastpm}). }
 \label{fig:8_robusness_tests}
\end{figure}

We proceed similarly with the non-Gaussian simulation. Starting from our fiducial choice of $k_{\rm max} = 0.1  \;  h\, {\rm Mpc}^{-1}$, we test both the lower and higher values. Our results show that varying the $k_{\rm max}$  has a minimal impact on $p$, with all central values remaining within  $1\sigma$ when $k_{\rm max}$ is varied from $0.07$ to $0.15 \;  h\, {\rm Mpc}^{-1}$. This is shown in panel {(a)} of \autoref{fig:8_robusness_tests}. The error bars  shrink by approximately $\sim 20 \%$ throughout this range when considering a larger $k_{\rm max}$.  Furthermore, we performed a linear fit for $p$ as a function of mass (as described in  Section \ref{sec:p_vs_m}) and we find that the slope is consistent with zero. If we compute the weighted average of bins 0--9, we find that $p(k_{\rm max}=0.07 \, h\,{\rm Mpc}^{-1}) = 0.933 \pm 0.013$ and $p(k_{\rm max}=0.15 \, h\,{\rm Mpc}^{-1}) = 0.947 \pm 0.012$, in agreement with our previous findings for our fiducial choice of $k_{\rm max}$.

Going beyond the purely linear modes may bias the constraints on $b_1$ (since those modes are not properly modelled) and hence the constraints on \bp and $p$. We therefore set the cutoff at  $k_{\rm max} = 0.1  \;  h\, {\rm Mpc}^{-1}$. On the one hand, we have shown that the results are consistent even if this value is varied by $\sim 50\%$. On the other hand, this is a standard choice, and it is more conservative than continuing to push to higher $k$, where we might have to deal with non-linear terms. 

\subsection{Power spectrum variance from \textsc{FastPM}}\label{sec:bphi_vs_fastpm}
An accurate estimation of the variance of the halo power spectrum, $\sigma(P(k))$, is critical when we perform parameter estimation, given that this quantity enters the computation of the likelihood. However, we know that the standard approach of \autoref{eq:pk_Gaussian_variance} overestimates the variance of our simulations as they have fixed initial conditions. In Section \ref{sec:variance_estimation}, we discuss how we estimated  $\sigma(P(k))$ from the \textsc{FastPM} mocks, which take into account the effect of the fixed ICs. We show how, by fitting $\sigma(P(k))/P(k)$ to \autoref{eq:sigpk_over_pk}, we can even extrapolate to masses where we are not able to generate the `equivalent' mass bins to the ones from the N-body simulation.

Here, we check the impact of considering different ways of computing the power spectrum variance on $p$. In particular, we consider the four different methods (see \S\ref{sec:variance_estimation}). First of all we consider the standard variances, $\sigma_S(P(k))$, computed from \autoref{eq:pk_Gaussian_variance} (blue symbols in panel (b) of \autoref{fig:8_robusness_tests}). Then we consider the direct \textsc{FastPM} variances, $sigma_F(P(k))$, obtained by computing directly the standard deviation of the power spectrum of the fast simulations (orange symbols). We also consider the best-fit of \textsc{FastPM} variances using \autoref{eq:sigpk_over_pk} (green symbols). And finally, we consider the extrapolation of \autoref{eq:sigpk_over_pk} to the mean halo mass of the bins (red symbols).

Our baseline analysis adopts the third method for bins 3-9, where we use the best fit of \autoref{eq:sigpk_over_pk} for each one, and the fourth method for bins 0-2, which are beyond the resolution limits of the \textsc{FastPM} mocks and hence, we cannot apply the third method directly.

We present the results of our comparison in panel {(b)} of  \autoref{fig:8_robusness_tests}, which shows the PNG bias parameter $p$ as a function of $M_{\rm halo}$ for the different methods of estimating the power spectrum variance described above. For bins 0--3, we do not have enough mass resolution with the  \textsc{FastPM} mocks, so only results assuming the standard variances and the extrapolation of $\sigma_F(P(k))$ are shown. Regardless of the method used to estimate the variance, we recover the same results within $1\sigma$. Moreover, the \textsc{FastPM} approach is more accurate than the standard approach as it considers the fixed initial conditions in our simulations. Additionally, the best fit of $\sigma_F(P(k))/P(k)$ provides almost the same results compared to using the extrapolation method. These results are consistent with using the variances directly obtained from \textsc{FastPM} mocks, even considering their small error bars. While we cannot test and compare our fitting procedure outside the mass domain of the \textsc{FastPM} mocks,  we remark that our results are still consistent within $1\sigma$ with the standard approach for all mass bins. 

We now consider the fit to mass bins 0-9. We performed a linear fit on $p$ and found that it is consistent with being a constant, confirming the results of Section \ref{sec:p_vs_m}. Furthermore, the weighted average yields $p(\sigma_{S+M}) = 0.950\pm0.056$ and $p({\rm direct\, \textsc{FastPM}}\,\sigma_{F+M}) = 0.937\pm0.021$, which agrees within $1\sigma$ with our fiducial choice.


\section{Constraining \fnl with priors on $p$}\label{sec:priors}

In this section, we study how setting a prior in $p$ (or equivalently, on \bp) affects the constraints on \fnl. This is because, with the scale-dependent bias, we are only sensitive to the product $b_\phi f_{\rm NL}$.

Here, we assume a Gaussian prior for $p$ centred on $p_{\rm prior}$ and with a width $\sigma(p)$. To implement the priors in our MCMC procedure (see Section \ref{sec:choices} for details), we modify the Gaussian likelihood by introducing this a second term in $\chi^2$:

\begin{equation}
    \chi^2 = \sum^{N_{k}}_{i=0} \frac{\left(P_{\rm halos}(k_i) - P_{\rm model}(k_i,b_\phi f_{\rm NL},b_1)\right)^2}{\sigma^2(P(k_i))} + \frac{(p-p_{\rm prior})^2}{\sigma^2(p)}\; .
    \label{eq:chi2_prior}
\end{equation}

In Section \ref{sec:bphi_prior}, we demonstrate how increasing the prior width $\sigma(p)$ deteriorates the constraints in \fnl. In Section \ref{sec:forecast}, we show how using these priors can affect the measurement of \fnl in future cosmological surveys.

\subsection{Effect of the prior size on $\sigma(f_{\rm NL})$}\label{sec:bphi_prior}
\begin{figure}
 \includegraphics[width=\columnwidth]{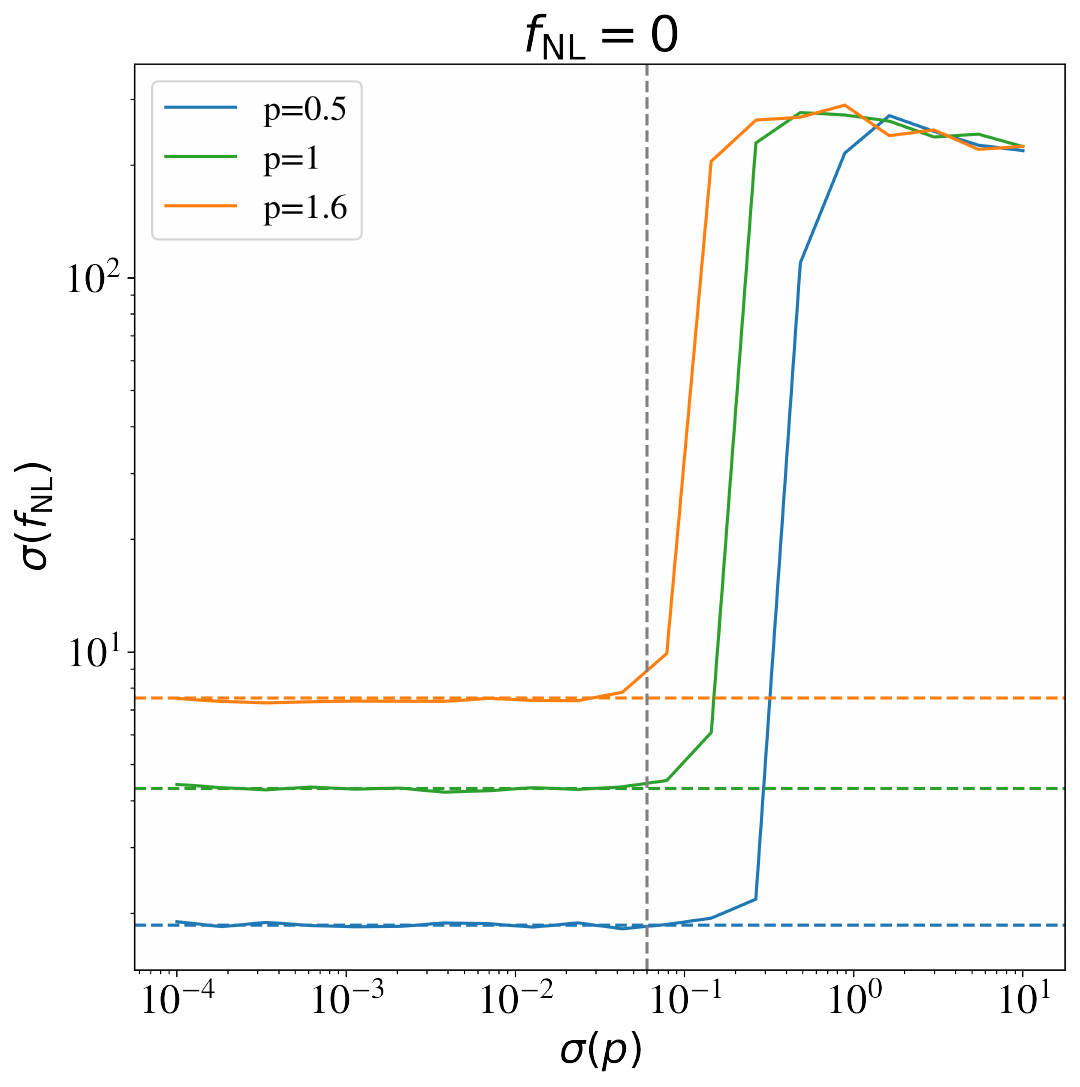}
 \caption{Statistical error in \fnl as a function of the size of the prior on $p$, for three values for $p$, shown in different colours, as indicated in the legend. The mock data vector for this figure is generated using the linear power spectrum computed with \textsc{CAMB} in \autoref{eq:power_spectrum_scale_dependent_bias} and fixing $b_1=1.4$ and $f_{\rm NL,\;fid}=0$. The horizontal dashed lines show the case where we performed the fits fixing $p$ in the model to its corresponding fiducial value. The dashed vertical  line displays the value of the error in $p$, obtained for halos with mass between $5\times 10^{11}$ and $1\times 10^{12} \, h^{-1} M_\odot$   (bin 4 in \autoref{tab:mass_bins}). }
 \label{fig:sigfnl}
\end{figure}

This section examines how the width of the prior for $p$ affects the determination of $f_{\rm NL}$. Nine mock halo power spectra are generated based on \autoref{eq:power_spectrum_scale_dependent_bias}. \textsc{CAMB} is used to compute the matter power spectrum in each case. Three $f_{\rm NL}$ values are considered: $0$, $-12$, and $50$, with $b_1$ set to $1.4$ for all cases. The choice of  $f_{\rm NL, \,fid\, 1} = 0$ is motivated by the current tightest constraint for $f_{\rm NL}$ by \citet{Planck_2018}; $f_{\rm NL, \,fid\, 1} = -12$ and $f_{\rm NL, \,fid\, 3} = 50$ are motivated by two recent constraints coming from galaxy clustering analysis \citep{Mueller_2021,Rezaie_2023}. The values of $b_{\phi}$ are derived in each case from \autoref{eq:bphi_p}, assuming that $p$ is $0.5$, $1$, and $1.6$. MCMC analyses are conducted to measure $f_{\rm NL}$ and its uncertainty $\sigma(f_{\rm NL})$ for each of the nine mocks.

We analyse the data in two ways. First, we fix $p$ to its fiducial value, assuming perfect knowledge of this quantity. Second, we allow $p$ to vary, introducing a Gaussian prior centred on the fiducial value of $p$ with a varying width. The idealised scenario of having a fixed $p$, has only two free parameters: $b_1$ and $f_{\rm NL}$. This scenario yields the most stringent constraints. Allowing $p$ to vary requires a prior on this parameter to counter the degeneracy between $p$ and $f_{\rm NL}$. This is further discussed below.

In \autoref{fig:sigfnl}, we show the uncertainty $\sigma(f_{\rm NL})$ for the setup described above as a function of the width of the prior in $p$, $\sigma(p)$, for the case of $f_{\rm NL}=0$. The analysis for the three fiducial $p$ values are shown in different colours.
The values of $\sigma(f_{\rm NL})$ from assuming fixed values for $p$ are shown as horizontal dashed lines. Those assuming a prior in $p$ are shown as solid lines.  

\autoref{fig:sigfnl} shows that constraints on $f_{\rm NL}$ are similar when considering either a fixed value of $p$ or a prior, as long as this prior is narrow enough, $\sigma(p) \lesssim 0.1$. Beyond this threshold, the constraints weaken significantly, increasing in $\sigma(f_{\rm NL})$. As we can see in \autoref{fig:sigfnl}, this threshold depends on the fiducial value of $p$. In fact, it also depends on $b_1$ and $b_{\phi}$. For a fixed $b_1$, the lower the $p$ (and hence the higher the $b_{\phi}$), the tighter the constraints on $f_{\rm NL}$. In \autoref{fig:sigfnl} we only show results for $f_{\rm NL}=0$, but similar trends are found for $f_{\rm NL}=-12$ and $50$.

\subsection{Forecast for a DESI-like survey}\label{sec:forecast}
\begin{table*}
\centering
\caption{Forecasts on \fnl  for halos hosting the different tracers targeted by DESI.}
\begin{tabular}{cccc|ccc |cc}
\cline{2-9}
                                &     \multicolumn{3}{c}{ELG hosts} &  \multicolumn{3}{c}{LRG hosts} &  \multicolumn{2}{c}{QSO hosts}\\\hline
                                &  $p=1$          & $p_{\rm fid}=0.895$       &prior   & $ p=1$           & $p_{\rm fid}=1.11$     &prior    & $p=1$               &  prior\\\hline
$f_{\rm NL}^{\rm fid} = 0$   & $-0.3\pm 3.4$   &$0.0\pm2.9$      & $-0.2\pm2.9$       & $-0.2\pm 3.0$    &$-0.3\pm3.1$  & $-0.5\pm3.4$  & $-0.3\pm 3.3$       & $-0.5 \pm 3.3$\\ \hline
$f_{\rm NL}^{\rm fid} = -12$ & $-15.1\pm3.3$   &$-12.5\pm2.7$    & $-12.7\pm2.9$      & $-11.4\pm2.3$    &$-12.4\pm2.3$ & $-13.0\pm3.1$ & $-11.9 \pm 3.7$     & $-12.1 \pm 3.4$\\ \hline
$f_{\rm NL}^{\rm fid} = 50$  & $60.4\pm5.5$    &$50.0\pm4.5$     & $50.3\pm5.7$       & $46.1\pm4.8$     &$49.7\pm4.9$  & $51.5\pm9.1$  & $50.8 \pm 3.5$      & $50.3 \pm 3.8$\\ \hline
\end{tabular}
\tablefoot{In the three sections of the table, we fix the fiducial values of $p$ to the values we find in \autoref{tab:mass_bins} for the halos with masses similar to those that host ELGs, LRGs and QSOs in \citet{Yuan_2023} (bin 4, bin 8, and bin 6) respectively. We also consider three fiducial \fnl values, as listed in each of the three rows. In  the first for each tracer, we assume $p=1$ and we fix \bp using \autoref{eq:bphi_p}. In the second column, labeled as $p_{\rm fid}$, we conduct the fitting in the same way, but we fix $p$ to the fiducial value for each sample. Finally, in the prior columns, we use our results for $p$ to conduct the fits with $p= 0.895 \pm 0.055$, $p = 1.11 \pm 0.27$, and $p = 0.973 \pm 0.077$ for the ELG, LRG, and QSO-like samples, respectively.}

\label{tab:mcmc_forecast}
\end{table*}

We go on to study the impact on the determination of \fnl in the context of current galaxy surveys when assuming either the universality relation or considering priors for $p$ (or \bp). We base our forecast on the clustering of the typical halos hosting cosmological tracers.

For this forecast, we assume survey properties similar to DESI, whose footprint will cover an area in the sky of $\sim14\,000 \deg^2$ \citep{DESI_2023}. We consider the three cosmological tracers targeted by the DESI dark-time programme: luminous red galaxies (LRGs), emission-line galaxies (ELGs) and quasars (QSOs). For our analysis, we use the results given in \citet{Jiaxi_2023} for the DESI early data release (EDR), as they are based the authors' analysis on the Gaussian \textsc{UNIT} simulation. \citeauthor{Jiaxi_2023} found that DESI EDR LRGs, ELGs, and QSOs populate halos with typical masses of $\log_{10}(M_{\rm halo,\, LRG}) = 13.16$, $\log_{10}(M_{\rm halo,\, ELG}) = 11.90$, and $\log_{10}(M_{\rm halo,\, QSO}) = 12.66$, respectively. These masses correspond to our halo mass bins $8$, $4$, and $6$, respectively (see \autoref{tab:mass_bins}).  Additionally, each tracer is restricted to a certain range of redshifts. The redshift range for LRGs is from $z=0.4$ to $z=1.1$, with an effective redshift of $z_{\rm eff\, LRG}=0.814$. For ELGs, the range goes from $z=0.8$ to $z=1.6$ with an effective redshift of $z_{\rm eff\, ELG}=1.202$. And for QSOs, we have considered the range from $z=0.8$ to $z=2.5$ with an effective redshift of $z_{\rm eff\,QSO}=1.741$. We computed  each tracer's total volume, assuming a spherical shell between the redshift ranges and the entire DESI footprint. In this way, we get $V_{\rm tot, \, LRG} = 19.45\, h^{-1}{\rm Gpc}$, $V_{\rm tot, \, ELG} = 34.37 \, h^{-1}{\rm Gpc}$, and $V_{\rm tot, \, QSO} = 64.94\, h^{-1}{\rm Gpc}$. The number densities \citeauthor{Jiaxi_2023} found for the considered redshift range  are $n_{\rm LRG} = 5.50 \times 10^{-4}\; h^{3}\, {\rm Mpc}^{-3}$, $n_{\rm ELG} = 7.26\times 10^{-4}\; h^{3}\, {\rm Mpc}^{-3}$ and $n_{\rm QSO} = 2.4 \times 10^{-5}\; h^{3}\, {\rm Mpc}^{-3}$. For the linear bias we assume $b_{1,\;{\rm LRG }}= 1.2/D(z)$, $b_{1,\;{\rm ELG }}= 0.84/D(z)$, and $b_{1,\;{\rm QSO}}= 1.2/D(z)$, where $D(z)$ is the growth factor \citep{DESI_2016,DESI_2023}. At the effective redshifts of interest, this leads to $b_{1,\;{\rm ELG }}= 1.503 $ , $b_{1,\;{\rm LRG }}= 2.568$, and $b_{1,\;{\rm QSO }}= 2.628$. 

We generated the mock halo power spectrum using \textsc{CAMB} in  \autoref{eq:power_spectrum_scale_dependent_bias}. As described in the previous section, here we consider three fiducial values for \fnl:  $f_{\rm NL, \,fid\, 1} = 0$, $f_{\rm NL, \,fid\, 2} = -12$, and $f_{\rm NL, \,fid\, 2} = 50$. Regarding \bp, we assume it follows the form given by \autoref{eq:bphi_p}. We use the values of $p$ measured in this work for the mass bin corresponding to each tracer\footnote{This is an assumption for the forecast in order to understand and highlight the impact of priors that \textsc{PNG-UNITsims} is able to set. However, we expect that redshift evolution and galaxy formation could modify the value of $p$, and we expect to tackle these in detail in follow-up studies.}: $p_{\rm LRG} = 1.11 \pm 0.27$, $p_{\rm ELG} = 0.895 \pm 0.055$ and $p_{\rm QSO} = 0.973 \pm 0.077$. The minimum wavenumber considered for the power spectrum is given by the fundamental wavenumber, $k_{\rm min,\, tracer} = 2\pi/V^{1/3}_{\rm tot,\, tracer}$. The corresponding covariance is calculated using \autoref{eq:pk_Gaussian_variance}, considering the survey parameters described above for each tracer.

We used an MCMC fitting procedure to determine the impact on the measurement of \fnl from assuming a prior for $p$ or not. We employed the same methodology as for our main results. In this analysis, all the cosmological parameters are fixed, except for  $b_1$, \fnl, and \bp.  When fixing \bp to the universality relation value, we defined the value of $\chi^2$  using \autoref{eq:chi2}. When considering $p$ as a free parameter, we incorporated a Gaussian prior centred on $p_{\rm LRG}$, $p_{\rm ELG}$ and $p_{\rm QSO}$ for each case value and a width given by our errors. In this case, $\chi^2$ is defined by \autoref{eq:chi2_prior}.

In \autoref{tab:mcmc_forecast}, we summarise our findings. Fixing $p$ to the universality relation induce a bias of $1.9 \sigma$ for $f_{\rm NL\; fid}=50$ and $0.9\sigma$ for $f_{\rm NL\; fid}=-12$ for ELGs. These biases arise because the halos hosting ELGs have a value for $p$ below one by $\sim 10\%$. For LRGs, we also find a $0.8 \sigma$ difference for the case with $f_{\rm NL\; fid}=50$. Finally, for the QSO hosts, we correctly recovered the fiducial values of \fnl given that the value of $p$ for this sample is consistent with $p=1$. 

When we fix $p=p_{\rm fid}$, we recovered unbiased results for \fnl as expected. Nevertheless, the errors in \fnl increase or decrease if $p_{\rm fid}$ is higher or lower compared to the case with $p=1$. These are the cases of the ELG hosts and LRG hosts, respectively. This variation in $\sigma(f_{\rm NL})$ is because the larger \bp is, the tighter the constraints on \fnl are, as we are only sensitive to the product of both quantities. We found negligible differences for the QSO hosts.

Finally, we recovered unbiased results for the three tracers when using our priors while, at the same time, the constraining power is comparable to fixing $p$ to its fiducial value. In the case of the ELG hosts, our constraints are equal for $f_{\rm NL}^{\rm fid}=0$, deteriorating by $7\%$ and a $ 27\%$ for $f_{\rm NL}^{\rm fid}=-12$, and $f_{\rm NL}^{\rm fid}=50$, respectively. For the LRG hosts, our priors perform worse, deteriorating the constraints by a $9\%$, a $34\%$  and a $85\%$ for $f_{\rm NL}^{\rm fid}=0$, $f_{\rm NL}^{\rm fid}=-12$, and $f_{\rm NL}^{\rm fid}=50$, respectively. Nevertheless, we are still able to constrain \fnl  without recovering the degeneration between this parameter and $p$. Finally, our priors for the QSO hosts are comparable to fixing $p=1$, providing even tighter constraints for $f_{\rm NL}^{\rm fid}=-12$.


\section{Summary and conclusions}\label{sec:conclusions}

Detecting primordial non-Gaussianities (PNG) would have a major impact on how we understand inflation. The key to measuring or accurately constraining PNG in the near future is to understand the scale-dependent bias parameter, \bp, and how it varies for different cosmological tracers.

In this paper, we study the scale-dependent bias \bp for dark matter halos with masses from $2\times 10^{10} \, h^{-1} M_\odot$ to $5\times 10^{14} \, h^{-1} M_\odot$. To achieve this, we have developed \textsc{PNG-UNIT}, a new full N-body simulation with local non-Gaussian initial conditions $(f_{\rm NL} = 100)$. \textsc{PNG-UNIT} is the largest simulation to date that incorporates non-Gaussian initial conditions. The simulation has $4096^3$ dark matter particles in a $1 \; (h^{-1}\,{\rm Gpc})^3$  box, which was run using the \textsc{L-Gadget2} code. The cosmology of the simulation is based on \citet{Planck_2015} (see \autoref{tab:cosmology}). \textsc{PNG-UNIT} assumes local PNG with an amplitude of $f_{\rm NL} = 100$. We have chosen a large value of \fnl to increase the signal from PNG while reducing the errors on \bp. We significantly reduced the error on \bp by using the method proposed in \citet{Avila_2022} for simulations with fixed-and-matched initial conditions. The amplitudes of the modes in the initial conditions are fixed to their expectation values, and the phases are matched to one of the realisations of the Gaussian \textsc{UNIT} simulations \citep{Chuang_2019}, following \citet{Avila_2022}.

In addition to the main simulation, \textsc{PNG-UNIT}, we also developed supporting simulations: a set of full N-body simulations, either with lower resolution or smaller volumes than \textsc{PNG-UNIT}, along with $100$ fast mocks,  half of which have $f_{\rm NL} = 100$ and the other half with $f_{\rm NL} = 0$, generated with the approximated method \textsc{FastPM}. We refer to the whole set of simulations as \pngsim. 

The \pngsim suite (see Section \ref{sec:png-unitsim}) is a unique laboratory for studying galaxy clustering models in the presence of local PNGs. The simulations in this suite have effective volumes comparable to current spectroscopic galaxy surveys. The reference simulation, \textsc{PNG-UNIT}, has a mass resolution high enough to accurately resolve  the dark matter halos expected to host the galaxies currently targeted by surveys such as DESI and EUCLID. Thus, the \pngsim suite is a powerful tool to develop and test analysis pipelines to measure \fnl with current spectroscopic galaxy surveys. 

In this first study of the \pngsim suite,  we were able to constrain the non-Gaussian bias parameter \bp using mass-selected samples of dark matter halos. We inferred the scale-dependent bias from the halo power spectrum measured for the Gaussian and the non-Gaussian simulations. From the power spectra, we inferred $b_1$ and $b_\phi f_{\rm NL}$ and correlated the measurements in both simulations Section \ref{sec:results}.
Moreover, given that we know the input values of \fnl, we can derive \bp or $p$. Our findings can be summarised as follows:
\begin{itemize}
    \item  We measured $b_\phi f_{\rm NL}$ using the fix-and-match technique (\S\ref{sec:matching}) by combining the information of the Gaussian and the non-Gaussian simulation. This technique  allows us to cancel out much of the cosmic variance (\S\ref{sec:bphifnl_vs_b1}).
    \item By assuming the universality relation, we recovered the input values of \fnl for our simulations within $1\sigma$ for halos between $1\times 10^{12}\, h^{-1} M_\odot$ and $1\times10^{14} \, h^{-1} M_\odot$. The power spectra are very noisy for higher halo masses due to their low number densities. As the variance estimates are not robust for these massive halos, we have not included them in the rest of our analyses. 
    \item For halos with masses between $5\times 10^{10}\, h^{-1} M_\odot$ and $1\times10^{12} \, h^{-1} M_\odot$ we find deviations from the input value of $f_{\rm NL}=100$, which reaches a significance of $4\sigma$ for halos between $1\times 10^{11}\, h^{-1} M_\odot$ and $2\times10^{11} \, h^{-1} M_\odot$, suggesting that the value of $p$ assumed ($p=1$) does not accurately described the clustering of these halos.
    \item We measured $p$ by fixing \fnl to the input values. For halos with masses higher than $1\times 10^{12}\, h^{-1} M_\odot$ and  between $2\times10^{10} \, h^{-1} M_\odot$ and $5\times10^{10} \, h^{-1} M_\odot$, we find $p$ to be consistent with the universality relation. For halos with masses between $5\times10^{10} \, h^{-1} M_\odot$ and $1\times10^{12} \, h^{-1} M_\odot$, we find $p$ to be $\sim 10\%$ below the universality relation, with a significance between $1.5 \sigma$ and $3.1 \sigma$.
    \item Combining the information of $b_\phi f_{\rm NL}$ from mass bins 0--9 ($M_{\rm halo}$ between $2\times 10^{10}\, h^{-1} M_\odot$ and $5\times 10^{13}\, h^{-1} M_\odot$) we find a preferred value of $p = 0.955 \pm 0.013 \, (68\% \,{\rm c.l.})$. This result is more than $3\sigma$ away from the universality relation (i.e. $p=1$).
   
\end{itemize}

We assured the validity of our results for $M_{\rm halo} \gtrsim 2 \times 10^{10} \, h^{-1} M_\odot$ by checking that the convergence with the LR-UNIT simulations ($2048^3$) occur at $M_{\rm halo}\sim20 m_{\rm part}$. Moreover, we also made a comparison with the separate universe technique, but the priors reported by the \textsc{PNG-UNIT} simulation are more robust. For the set-up considered here, we find that the separate universe tends to underestimate the uncertainty reported on \bp. By looking at the scatter and lack of convergence of the separate universe results, we also conclude that the uncertainty associated with those \bp measurements would be larger than the one found by \pngsim.

We explored how this affects variations on the mass definition used for the mass cuts to the measurements of $p$, finding consistency for all mass definitions except for $M_{500c}$(\S\ref{sec:bphi_vs_mdef}). We argue that \bp might not depend only on the halo mass through $b_1$ but also on other parameters. We leave a more detailed exploration of these dependencies for future work.

We have also tested the robustness of the above results to variations in the choice of the scale cuts (\S\ref{sec:bphi_vs_kmax}) and the way we compute the variances of the power spectrum (\S\ref{sec:bphi_vs_fastpm}). None of the above is shown to have a significant impact. 

Finally, we construct priors on the scale-dependent bias for DESI galaxies based on the simplified assumption that those tracers can be identified directly by the mass of their host halo at $z=1$. We show that these priors are:
\begin{enumerate}
    \item requiered to get unbiased constraints on \fnl (compared to the universality relation that may bias the results, up to $2\sigma$ for the setup considered);
    \item sufficiently tight to allow us to constrain \fnl with DESI at a comparable precision with respect to cases where $p$ is fixed to the fiducial value.
\end{enumerate}

All these results show the capability of the \pngsim suite as a valuable tool for interpreting observations with DESI and EUCLID. Specifically, we address the issue of putting priors on \bp. While we have analysed how the halo power spectrum can be used to constrain the scale-dependent bias in this work, the \pngsim suite is a unique tool to test and develop alternative methods for constraining PNG with cosmological surveys. In the future, we plan to include galaxies in our simulations and consider other clustering statistics such as the configuration space correlation functions, higher order statistics including the bispectrum or trispectrum, and one-point statistics such as counts-in-cells, density peaks, and galaxy cluster counts.


\section*{Data Availability}
The halo catalogues used in this paper are available through the \textsc{UNITsim} website: \url{http://www.unitsims.org/}
The other data products will be shared under email request to the corresponding author.


\begin{acknowledgements}

We thank V. Yankelevich, O. Hahn, M. Manera, and R. Scoccimarro for the valuable discussions on how to solve some early problems with initial conditions. We thank the Benasque science center for fostering useful discussions regarding this work. 
This work has been supported by Ministerio de Ciencia e Innovaci\'{o}n (MICINN) under the following research grants:
PID2021-122603NB-C21 (AGA, VGP, GY and AK), 
PID2021-123012NB-C41 (SA) and PID2021-123012NB-C43 (JGB). AK further thanks Embrace for `gravity'.
SA and VGP have been or are supported by the Atracci\'{o}n de Talento Contract no. 2019-T1/TIC-12702 granted by the Comunidad de Madrid in Spain. JGB has also been supported by the Centro de Excelencia Severo Ochoa Program CEX2020-001007-S at IFT. 
IFAE is partially funded by the CERCA program of the Generalitat de Catalunya.
The \textsc{PNG-UNIT} simulation has been run thanks to the computer resources at MareNostrum granted by the Red Española de Supercomputación  and the technical support provided by Barcelona Supercomputing Center (RES-AECT-2021-3-0004, RES-AECT-1-0007, RES-AECT-2022-3-0030, RES-AECT-2023-1-0014, and RES-AECT-2023-2-0006.).
\end{acknowledgements}

%
%
\bibliographystyle{aa}
\bibliography{my_biblio}



\end{document}